\documentclass[secnumarabic,amssymb, nobibnotes, aps, prf,superscriptaddress]{revtex4-2}
\usepackage{titlesec}
\usepackage{comment}

\usepackage[margin=0.7in]{geometry} 

\usepackage{graphicx, float}
\usepackage{dcolumn}
\usepackage{bm}

\usepackage{amsmath,amssymb,amsfonts, subeqnarray}
\usepackage{fancyhdr}
\usepackage{lipsum}
\usepackage{subcaption}
\usepackage{multirow}
\usepackage{tikz,comment} 

\usepackage[mathlines]{lineno}
\usepackage{etoolbox} 

\titleformat{\section}[hang]{\normalfont\Large\bfseries}{\thesection}{1em}{}

\titlespacing{\section}{0pt}{.5 \baselineskip}{2pt}

\newcommand*\linenomathpatch[1]{%
	\cspreto{#1}{\linenomath}%
	\cspreto{#1*}{\linenomath}%
	\csappto{end#1}{\endlinenomath}%
	\csappto{end#1*}{\endlinenomath}%
}
\newcommand*\linenomathpatchAMS[1]{%
	\cspreto{#1}{\linenomathAMS}%
	\cspreto{#1*}{\linenomathAMS}%
	\csappto{end#1}{\endlinenomath}%
	\csappto{end#1*}{\endlinenomath}%
}

\expandafter\ifx\linenomath\linenomathWithnumbers
\let\linenomathAMS\linenomathWithnumbers
\patchcmd\linenomathAMS{\advance\postdisplaypenalty\linenopenalty}{}{}{}
\else
\let\linenomathAMS\linenomathNonumbers
\fi

\linenomathpatch{equation}
\linenomathpatchAMS{gather}
\linenomathpatchAMS{multline}
\linenomathpatchAMS{align}
\linenomathpatchAMS{alignat}
\linenomathpatchAMS{flalign}
\linenomathpatch{subeqnarray}

\newcommand{\bs}{\boldsymbol}
\newcommand{\dbs}[1]{\dot{\boldsymbol{#1}}}
\newcommand{\tn}{\wideoverline{\bs{\nabla}}}
\newcommand{\on}{\bs{\nabla}}
\newcommand{\wb}[1]{\wideoverline{\bs{{#1}}}}
\newcommand{\bbb}[1]{\overline{\bs{#1}}}
\newcommand{\obb}[1]{\hat{\overline{#1}}}
\newcommand{\oa}[1]{\wideoverline{\mathcal{#1}}}
\newcommand{\rb}[1]{\bs{\mathcal{#1}}}
\newcommand{\pf}[2]{\frac{\partial #1}{\partial #2}}
\newcommand{\pfw}[2]{\frac{\partial \wideoverline{#1}}{\partial \wideoverline{#2}}}
\newcommand{\pfm}[3]{\frac{\partial^#1 #2}{\partial #3^#1}}
\newcommand{\pft}[2]{\frac{\partial^2 #1}{\partial #2^2}}
\newcommand{\pfh}[2]{\frac{\partial^3 #1}{\partial #2^3}}
\newcommand{\df}[2]{\frac{\mathrm{d} #1}{\mathrm{d}  #2}}
\newcommand{\dfm}[3]{\frac{\mathrm{d} ^#1 #2}{\mathrm{d}  #3^#1}}
\newcommand{\dft}[2]{\frac{\mathrm{d} ^2 #1}{\mathrm{d}  #2^2}}
\newcommand{\md}{\mathrm{d}}
\newcommand{\f}{p^\text{off}}
\newcommand{\n}{p^\text{on}}

\newcommand{\fw}{P^\text{off}}
\newcommand{\nw}{P^\text{on}}

\newcommand{\m}{p^\text{m}}
\newcommand{\pd}{p^\text{d}}
\newcommand{\mw}{P^\text{m}}
\newcommand{\dw}{P^\text{d}}
\newcommand{\ab}[1]{\langle #1 \rangle}
\newcommand{\abp}[1]{\langle #1 \rangle_{\bs{p}}}
\newcommand{\abpp}[1]{\langle #1 \rangle_{\bs{pp}}}
\newcommand{\re}{\rho_\text{eq}}
\newcommand{\R}{R_\text{eq}}
\newcommand{\w}[1]{\wideoverline{#1}}
\newcommand{\pe}{\mathrm{Pe}}
\newcommand{\nb}{\bs{\nabla}_{\bs{x}}}
\newcommand{\np}{\bs{\nabla}_{\bs{p}}}
\newcommand{\Dt}[1]{\frac{\mathrm{D} #1}{\mathrm{D} t}}

\begin{document}

\title{Dynamic Flow Control Through Active Matter Programming Language}

\author{Fan Yang}
\email[E-mail: ]{fy2@caltech.edu}
\affiliation{Division of Biology and Biological Engineering, California Institute of Technology, Pasadena, CA, USA}

\author{Shichen Liu}
\affiliation{Division of Biology and Biological Engineering, California Institute of Technology, Pasadena, CA, USA}

\author{Heun Jin Lee}
\affiliation{Department of Applied Physics, California Institute of Technology, Pasadena, CA, USA}

\author{Rob Phillips}

\affiliation{Division of Biology and Biological Engineering, California Institute of Technology, Pasadena, CA, USA}

\affiliation{Department of Applied Physics, California Institute of Technology, Pasadena, CA, USA}

\author{Matt Thomson}
\email[E-mail: ]{mthomson@caltech.edu}

\affiliation{Division of Biology and Biological Engineering, California Institute of Technology, Pasadena, CA, USA}

\begin{abstract}
	Cells control fluid flows with a spatial and temporal precision that far exceeds the capabilities of current microfluidic technologies. Cells achieve this superior spatio-temporal control by harnessing dynamic networks of cytoskeleton and motor proteins. Thus, engineering systems to mimic cytoskeletal protein networks could lead to the development of a new, active-matter-powered microfluidic device with improved performance over the existing technologies. However, reconstituted motor-microtubule systems conventionally generate chaotic  flows  and cannot perform useful tasks. Here, we develop an all-optical platform for programming flow fields for transport, separation and mixing of cells and particles using networks of microtubules and motor proteins reconstituted \textit{in vitro}. We employ mathematical modeling for design optimization, which enables the construction of flow fields that achieves micron-scale transport. We use the platform to demonstrate that active-matter-generated flow fields can probe the extensional rheology of polymers, such as DNA,  achieve transport and mixing of beads and human cells, and isolation of human cell clusters. Our findings provide a bio-inspired pathway for programmatically engineering dynamic micron-scale flows and demonstrate the vast potential of active matter systems as an engineering technology.
	
\end{abstract}
\maketitle

The control of micron-scale transport and fluid flow is a foundation of modern technology including synthetic chemistry, DNA sequencing and single-cell genomics \cite{stone04, whitesides06}. However, current microfluidic paradigms
usually require soft lithography in PDMS that sculpts the geometry of channels to control flow. The channels are coupled to macroscopic pumps and valves to drive flow. In contrast, active-force-generating biological materials provide an alternate technology paradigm for driving and controlling micron-scale flows. Biological systems generate fluid flows through active, energy consuming networks of cytoskeletal filament and motor proteins \cite{gilpin17}. Active networks consume energy at microscope length scales through ATP/GTP hydrolysis by individual motor and filament proteins. Self-organization and protein-based regulation enable cells to control networks to induce and modulate fluid flows in processes like cytoplasmic streaming and foraging \cite{kamiya81,glotzer97,forrest03, needleman19, stein21}. Cytoskeletal active matter could provide a novel technology platform for dynamic control of fluids in technology applications as decades of research have demonstrated that purified motor-filament proteins can generate micron-scale flows in solution \cite{suzuki17, wu17, stein21}. Furthermore, the dynamics of active matter can be controlled with light proving a potential platform for optically programmable micron-scale transport \cite{ross19}. Active matter flow control could enable programmable execution of dynamic micron-scale tasks including transport, separation, sorting and mixing. 


However, a fundamental challenge is that active fluids are historically thought to be difficult to control and harness for applications, due to  the non-linear coupled dynamics between active matter and fluids \cite{wu17}. Specifically previous work \cite{wu17} demonstrates that active fluids can exhibit a phenomenon known as ``active turbulence" where generated flow fields exhibit vortices and other transient structures that have similarities to macroscopic turbulence. While active flows can be controlled through fabrication of microfluidic chambers with designed boundaries \cite{wu17}, prefabricated geometries inherently limit the application of geometrically controlled active fluids and are not able to take advantage of or generate the dynamic spatio-temporal modulation of flows induced by active matter in biological systems \cite{gilpin17,mathijssen19}. Fundamentally, conceptual and theoretical paradigms that enable the control of active fluids could provide insight into the physical principles governing the dynamics of force-generating active fluids. 

In this paper, we develop a spatio-temporally flexible programming paradigm for modular design and construction of micron-scale flow fields using light-controlled biological active matter. We use an engineered system in which motor protein activity is modulated by light \cite{ross19}. By composing ``primitive" flow fields generated by a single rectangular light bar, we are able to generate flow fields that enable transport, stretching and separation of micron-scale particles and cells. Fundamentally, our  paradigm is based on the addition, mathematically known as superposition, of flow fields generated from rectangular bars.  Conceptually, low-Reynolds-number (low-Re) flows are linear and so flow fields induced by multiple point forces generate a composite flow field that is a simple sum or superposition of the flow fields generated by the point forces individually. However, while generally at low Re, active fluids break the linearity due to inherent non-linearity within active materials and the transport of the active materials by a flow field, for which we derive a design principle to restore dynamic linearity to active fluids.

\begin{figure}[H]
	\includegraphics[width=1\linewidth,angle=0]{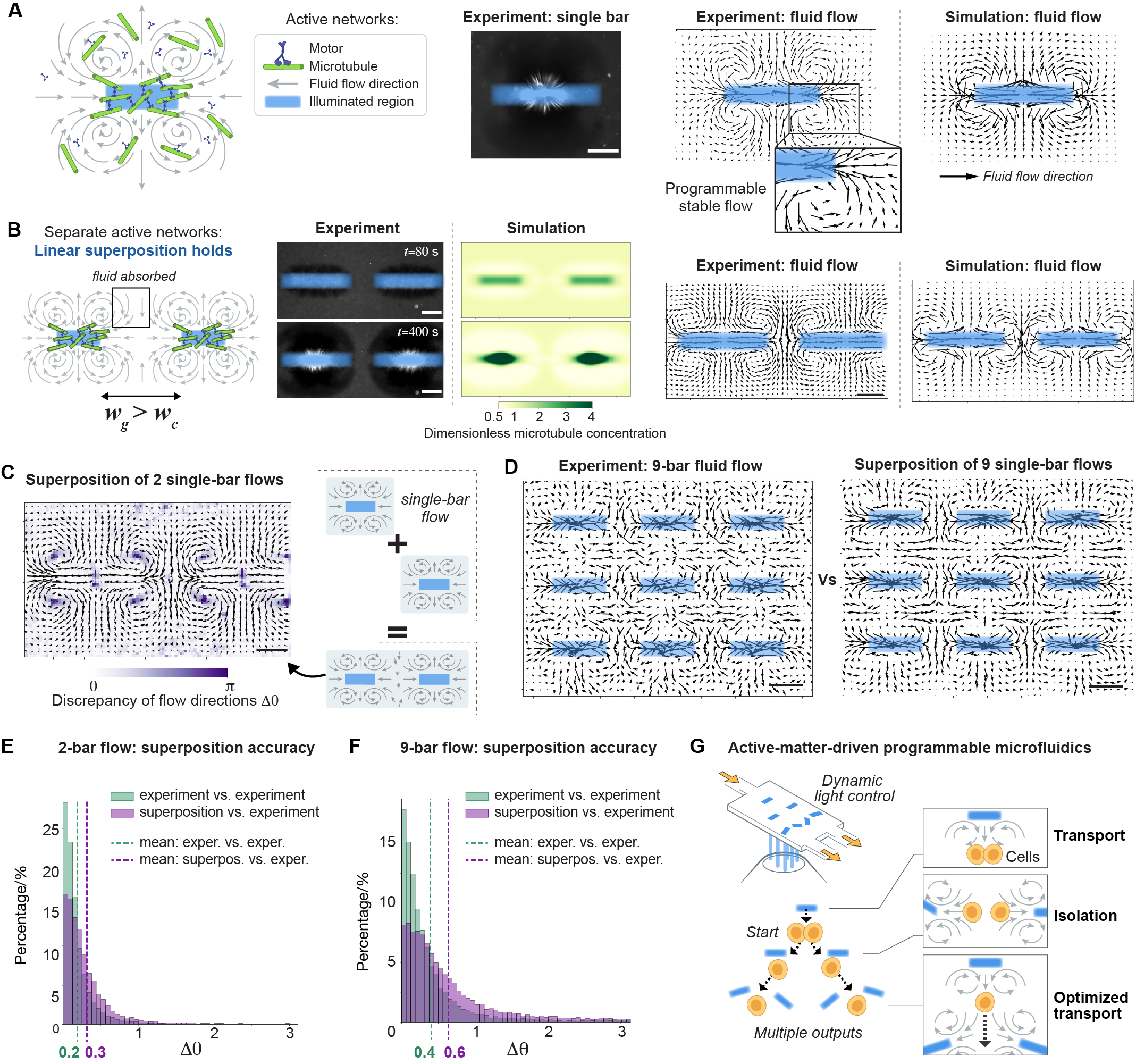}
	\caption{\textbf{Linear superposition quantitatively predicts fluid flow fields induced by optically-controlled active matter}. \textbf{A}, Under illumination, light-activatable motors can dimerize and crosslink microtubules into a network. The contractile active network generates flows in the surrounding fluid which is absorbed lengthwise into the light bar and pumped out widthwise. \textbf{B}, Above a critical spacing  $w_c$, the flow field generated by two light bars is a linear superposition of two single-bar flows.   \textbf{C}, Comparison of the measured two-bar flow field with superposition of two single-bar flows shows that the linear superposition holds quantitatively. The heatmap represents the angle $\Delta \theta \in [0, \pi]$ between measured and superposed flows. \textbf{D}, Measured 9-bar flow fields and superposition of 9 single-bar flows to assemble a 9-bar-array flow. \textbf{E} and \textbf{F},  Superposition provides high-accuracy flow field prediction for both 2-bar (\textbf{E}) and 9-bar (\textbf{F}) compositions, with errors comparable to variations in experimental replicates.  \textbf{G} Linear superposition enables construction of active-matter-driven programmable microfluidics. The sketch shows an conceptual application using assembly of light bars to transport and separate cells. All scale bars  are 100 $\mu$m and all  flow fields are time-averaged over 240 s. }
	\label{fig1}
\end{figure}


Through a combination of theory and experiment, we demonstrate that superposition can be recovered and applied to program flow fields using rectangular light patterns separated by a cut-off distance. We establish a quantitative continuum model for the coupled dynamics of active matter and background fluid, which is used in predictive programming design and optimization, and also uncovers the mechanism of the cut-off distance. Optimal control of active nematics has already been explored theoretically by solving a partial-differential-equation (PDE) system \cite{norton20}, whereas superposition can significantly simplify this  process by only using translation and rotation of simulated flow data generated by a single bar, which is  realized in experiments for particle transport. We also apply superposition-based flow programming to generate flow fields for extensional rheology of polymers and micron-scale manipulation tasks on primary human cells, such as separating an unconstrained cell cluster into individual cells {\itshape in situ}.   The advantages of our system are: there are no requirements on precise channel design and microfabrication, PDMS lithography or pressure-pump control; our system can generate local flows around objects of interest without disturbing other regions in the channel; different programming modules can be additively assembled for specific transport tasks, enabling streamlining operations and multi-tasking in a single channel; the system also allows us to move and control primary human cells, providing a novel platform for programmable manipulation of particles in biology and chemistry.

\section*{Flow programming through linear superposition of light bars} 


We specifically seek to develop a modular framework where we can compose a basic set of primitive light patterns, such as rectangular bars, to program microfluidics. Our inspiration is  Stokes flows, where  inertial effects can be neglected  and fluid dynamics can be described by a linear PDE known as the Stokes equation, $\mu \nabla^2 \bs{u}-\bs{\nabla}p=\bs{0}$, with $\bs{u}$ the flow velocity, $\mu$ the viscosity and $p$ the pressure. The amazing power of the dynamic linearity in Stokes flows is that, if we know the flow field generated by a single point source, then we can  compose points sources and the resulting flow field can be simply predicted through addition of flow fields generated by these point forces individually \cite{leal07}. Superposition provides a substantial simplification for predicting and programming flows through simple addition of point sources. However, in our system we are sculpting flows using active-matter-generated forces, and in general, linear superposition does not hold for active fluids due to their inherent non-linearity and apparent disorder \cite{marchetti13, wu17}.

We now introduce the design principle for linear superposition in active fluids with a set of force-generating agents.  Our experimental system consists of stabilized microtubules, and kinesin motor proteins that have been engineered to ``link" in the presence of blue light. In previous work \cite{ross19}, we have demonstrated that light induction generates contractile motor-filament networks that induce spontaneous fluid flows within the system. In the Supporting Information, we formulate a continuum model that can quantitatively predict dynamics of active matter and solvent flows. Our model is a three-phase complex fluids model, and the three phases are crosslinked microtubules, freely-moving microtubules and  solvent fluid. The crosslinked microtubules  are modeled as a viscoelastic gel, which self-contracts driven by its internal active stresses; the freely-moving microtubules are passive particles carried by both the gel and the solvent flow; the solvent flow is generated by the contraction of the active gel and balanced by the hydrodynamic resistance in the flow cell. 


In our model, the solvent flow is governed by the Stokes equation with  a  driving force applied by the motion of crosslinked microtubules:
\begin{equation}
 \gamma c (\bs{v}-\bs{u}) + \mu \nabla^2 \bs{u} -\bs{\nabla} p = 0,
\end{equation}
where  $c$ and $\bs{v}$ are the spatially varying concentration and velocity of the  crosslinked microtubules, respectively, and $\gamma$ is the drag coefficient between the microtubules and the solvent. In equation (1), $\gamma c (\bs{v}-\bs{u})$ acts as a field of point forces applied upon the fluid.  However, unlike the body-force-free Stokes equation, which is a purely linear, time-independent PDE, active fluids are transient and non-linear. Furthermore, the microtubule concentration field $c$ is also carried by the solvent flow $\bs{u}$ (Supplemental Information), so that the microtubule network at a position $\bs{x}_i$ experiences dynamics due to the long-range flows induced by microtubule network at positions $\bs{x}_j$ in the system. In general $c$ is therefore both time-dependent and also an implicit function of the ambient flow field, $\bs{u}$. Fundamentally, the problem is that microtubule networks activated at different locations within the system interact through  fluid flows, and flow-induced interactions lead to transport of the microtubule network giving rise to non-linear stresses and transport phenomena within the active fluids.  

We find that superposition can be restored in the model by restricting interactions among spatially isolated regions of light signals.  Mathematically, we show that the flow field $\bs{u}_i$ generated by a single network $i$ decays as a power law $x^{-3.5}$ with $x$ the distance to the bar center. When another network $j$ is placed beyond a cut-off distance such that $c_j\bs{u}_i = 0$ (Materials and Methods), superposition is recovered, and the flow field $\bs{u}$ generated by a system of spatially isolated light  patterns  can be predicted through simple linear superposition of individual forces in equation (1), that is, $\gamma c (\bs{v}-\bs{u}) = \gamma \sum_i c_i (\bs{v}_i-\bs{u}_i)$, where the subscript $i$ indicates the microtubule concentration, velocity and the solvent velocity induced by a single network $i$, in the absence of other networks (Materials and Methods). This tells us that to maintain linearity in a system with multiple active agents at low Re, the flow fields induced by each agent should decay to very small at the locations of other agents comparing to their self-generated velocities.  Therefore, we develop a modular programming strategy where we minimize long-range interactions by positioning the isolated light patterns far enough from each other so that the activated networks only weakly interact.

\section*{Linear superposition principle enables quantitative prediction of experimental flow fields} Consistent with our theory,  we find experimentally that the flow field generated by two rectangular light bars can be predicted by superposition when the bars are separated above a critical spacing. We use a rectangular light bar as the basic unit of programming design, which dynamically functions as a microfluidic pump: the active network absorbs fluid lengthwise and pumps it out widthwise generating four counter-rotating vortices (Fig. \ref{fig1}A). We test our principle for linear superposition with two light bars placed side by side. When their gap width $w_g$ exceeds a critical spacing $w_c$, the active networks self-contract within their respective illuminated regions (Fig. \ref{fig1}B). The resultant microtubule and flow fields (Fig. \ref{fig1}B) are, at least qualitatively, linear superposition of two single-bar fields (Fig. \ref{fig1}A).


We find that linear superposition of single-bar flows can quantitatively predict flow fields generated by multi-bar compositions. We additively assemble  the single-bar flow fields (Fig. 1A) to construct 2-bar (Fig. 1C) and 9-bar (Fig. 1D) fluid flows, and plot the discrepancy of flow directions with measured flows in Fig. \ref{fig1}E and Fig. \ref{fig1}F, respectively. There are mainly two sources of discrepancy: the errors induced by superposition and the  experimental variations from inherent thermal fluctuations. To distinguish them, we compare the distributions of discrepancy from superposition with the distributions of experimental variations (Fig. 1E and 1F), which show that the errors induced by superposition are comparable to the intrinsic thermal fluctuations. Similarly we also compare the discrepancy of flow magnitudes in superposition of 2-bar flows with experimental variations (Fig. S1, Supplemental Information). The mean fractional change of  magnitudes induced by superposition is $0.31$, only slightly above  the mean in experimental variations, which is $0.23$. The small superposition-induced errors  in both flow magnitudes and directions demonstrate that linear superposition can be achieved quantitatively in multi-bar compositions. Linear superposition is long thought to be impossible in active-matter systems, while in our experiments it is made possible by confining the active matter within the illuminated regions. Outside the light regions,  microtubules and monomer motors do not crosslink and dynamic linearity of Stokes flows still holds. Linear superposition is the foundation of constructing a modular programming language for microfluidic control (Fig. 1G). In our control strategy, only fluid flows outside the illuminated regions are utilized for transport tasks (Fig. 1G). 

\section*{Hydrodynamic interactions between active networks determine superposition length scales }

The linear superposition fails when the spacing between light bars are below a critical length. The active networks move outside the light regions and eventually merge (Fig. 2A). Our continuum model allows us to determine how the critical spacing originates from hydrodynamic interactions between active networks. The rectangle active networks absorb fluid flow along the long  axis ($x$-axis in Fig. 2B), and therefore, can attract neighboring networks through hydrodynamic interactions. The flow decaying with distance provides a mechanism for whether the two active networks will merge. In the 2-bar composition (Fig. 1B and 2A), each network generates flow that propagates to the neighboring network . The flow field generated by one network, therefore,  exerts force  on the neighboring network leading to flow-induced drift that is also counteracted by activity-induced self-contraction. When bars are placed close enough, within a critical length $w_c$, the flow generated by one network is sufficient to move the other network out of the illumination region against self-contraction (Fig. 2A). 

How flow decays with the distance to the bar center is essential in determining the critical spacing, which is found to follow a power law. The intensity of a flow field generated by a given rectangular network determines the network spacing required for superposition to hold.  We first conduct  simulations and experiments with two light bars of different sizes and similar aspect ratios (Fig. 2B).  The scaling of the flow magnitude is the same for both bars, which grows linearly and decays as $x^{-3.5}$ within and outside the illuminated region, respectively. The simulation results are in good agreement with experiments, except for a small region near the  edge of the larger light bar. The flow magnitude drops slightly in this region, which may suggest that a boundary layer is developed here when the light bar is large enough. The flow data for two sizes of bars collapse onto each other when the distance $x$ is normalized by the bar length $w$ (Fig. 2B), which demonstrates that $w$ is the length scale of flow fields. In practice, a spacing of half of the bar length $w/2$ is usually enough to avoid merging of networks as the flow already decays by more than 98\% at $x=3w/2$.    To test the generality of superposition, we also compute the effects of aspect ratios of light bars on the flow decay. Numerical results reveal that the flows outside the illuminated regions always decay in a power law, with the power ranging from -4 to -3.5, regardless of the aspect ratios (Fig. S3, Supplemental Information). This demonstrates that linear superposition can be generalized to different bar sizes and shapes. 

To test how the flow intensity affects the critical spacing, we calculate a phase diagram of two networks with different gap widths and flow intensities (Fig. 2C).  The flow intensity is tuned by a dimensionless parameter $\overline{\zeta}= c_0 \gamma h^2/ 12 \mu$, where $c_0$ is the typical microtubule concentration and $h$ is the height of the flow cell.  This parameter is found to be the most important dimensionless number governing the flow intensity in our model (Eqn S28, Supplemental Information). The physical meaning of $\overline{\zeta}$ is the ratio of the driving force, $c_0 \gamma$, and the hydrodynamic resistance in the flow cell, $12\mu/h^2$.   The phase  diagram (Fig. 2C) shows that  the critical spacing required for superposition increases with the flow intensity and thus the microtubule concentration. The reason is that as the flow intensity generated by network one increases, network two, its neighboring network, should be placed farther away. This ensures that the flow induced by network one decays to be no longer sufficient to drag network two outside the illuminated region.


\begin{figure}[t]
	\centering
	\includegraphics[width=1\linewidth,angle=0]{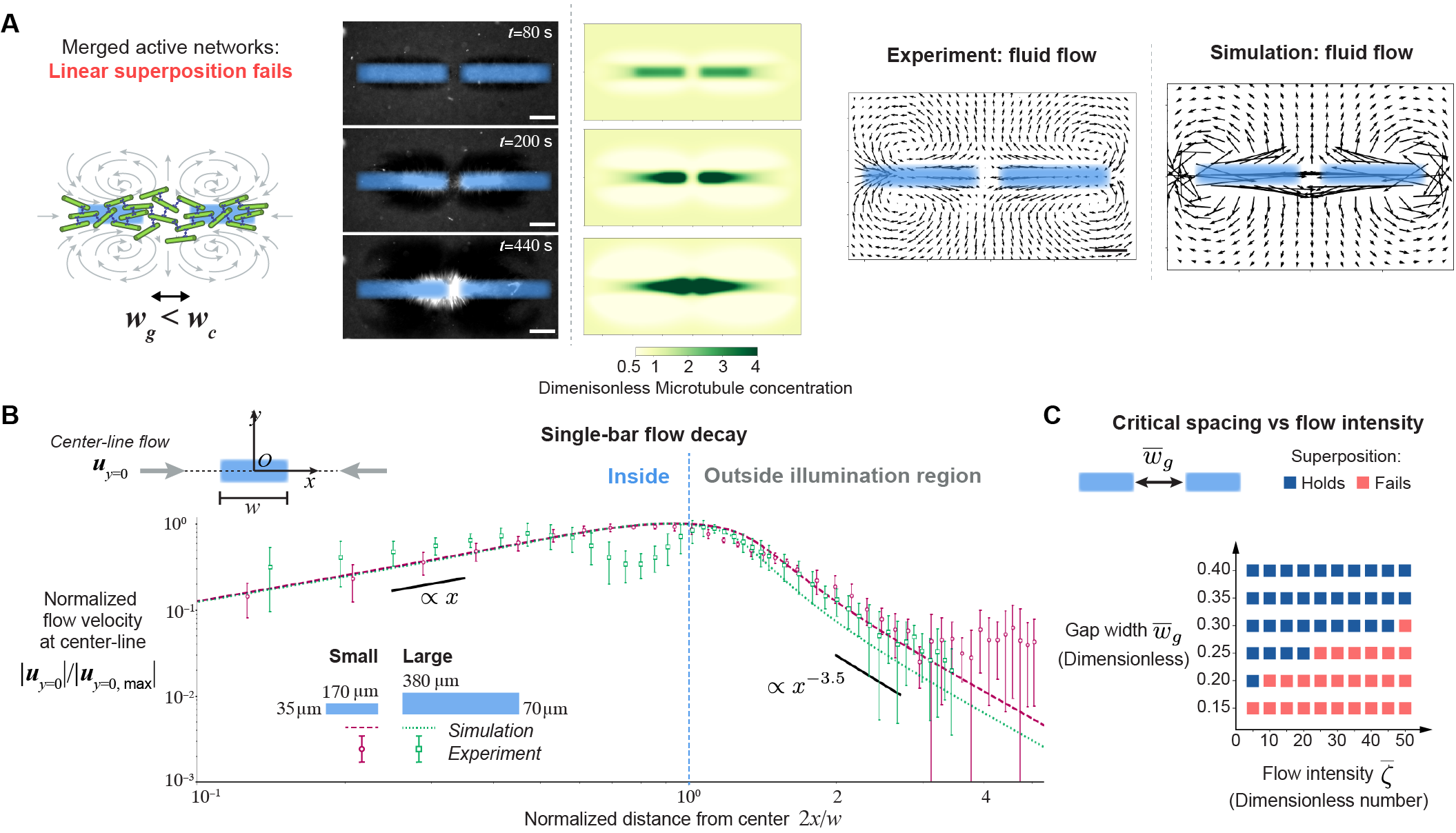}
	\caption{\textbf{Superposition length scale is set by hydrodynamic interactions between active networks}. \textbf{A}, Below the critical spacing, linear superposition of flow fields fails. The active networks move towards each other and eventually merge. \textbf{B}, The center-line flow profile induced by a single bar.  \textbf{C}, Simulated phase diagram of superposition shows that the critical spacing increases with the fluid flow intensity. All scale bars are 100 $\mu$m and  the flow fields are averaged over  $240$ s. } 
	\label{fig2}
\end{figure}

\section*{Design of multi-bar flow fields for transport and stretching with numerical optimization}

\begin{figure}[H]
	\centering
	\includegraphics[width=0.88\linewidth,angle=0]{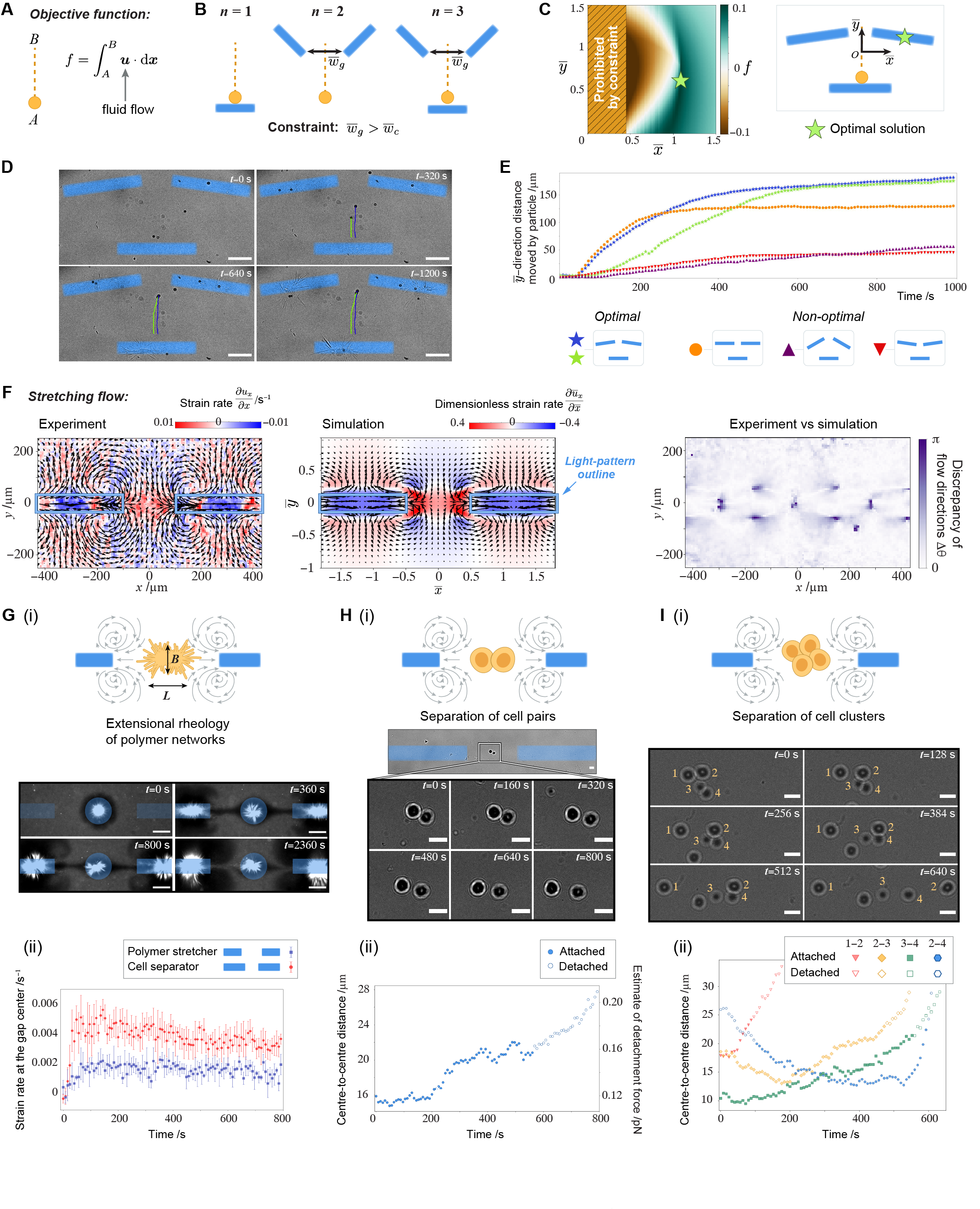}
	\caption{\textbf{Superposition provides a simple design principle for optimized transport of particles, microrheology of polymer and isolation of cells}. \textbf{A}, The design  goal, to move a particle from point A to B, and its corresponding objective function. \textbf{B}, The particle path AB should coincide with the axis of symmetry in the composition of light bars. \textbf{C}, The optimal  3-bar configuration is determined by the maximum value in the optimization landscape and displayed on the right. \textbf{D}, Experimental images of using the optimal light bar configuration (blue) to transport two particles.  \textbf{E},  Particle transport over time for optimal vs non-optimal designs. \textbf{F}, Measured and simulated stretching flows generated by two light bars, which agree quantitatively with each other. Both flow fields are averaged over $240$ s. \textbf{G}, (i) Experimental images of a pre-formed microtubule aster being stretched by the elongational flow, which is used to measure rheological properties of the aster; (ii), Measured flow strain rates over time in the polymer-stretching and cell-separating experiments. The strain rates are averaged over a $140 \times 50 \ \mu$m region at the gap center.  Error bars represent the standard deviations of at least four measurements. \textbf{H}, (i) Snapshots of detaching two cells; (ii), Centre-to-centre distance of the two cells and estimated detachment force over time.   \textbf{i}, (i) Snapshots of separation of a cell cluster in the stretching flow. (ii) Centre-to-centre distances of cells  over time. Due to constraints of image resolution, the time of detachment has an error of $\pm$30 s. The scale bars are 100 $\mu$m in \textbf{D} and \textbf{G}, and 20 $\mu$m in \textbf{H} and \textbf{I}.}
	\label{fig3}
	
\end{figure}

Superposition opens up a convenient route to design of flow fields using optimization via translation, rotation and summation of single-bar flow data. We study an optimization problem of using three light bars to transport particles along a line segment AB (Fig. \ref{fig3}A). The objective function $f$ is defined as the line integral of fluid flows along AB (Fig. \ref{fig3}A). To ensure the particle transport along a straight path, the line segment AB should be placed at the axis of symmetry of the bar configurations (Fig. \ref{fig3}B). Using simulated flow data, we can  plot the optimization landscape in Fig. \ref{fig3}C, where each point value represents a configuration of the three light bars (Supplemental Information).  The optimal solution corresponds to the maximum value in the optimization landscape, labeled by a star in Fig. \ref{fig3}C. We test this optimal solution in experiments by comparison with non-optimal three-bar configurations,  and find that the  optimal solution  transports particles both fastest and furthest (Fig. \ref{fig3}D and \ref{fig3}E). The results show that active matter programming language is capable of optimizing practical design using only linear transformation of single-bar flow fields.

\section*{Programming flows for microrheology and cell isolation} To show that active matter programming language can be flexibly applied and motivate new applications, we utilize  stretching flows induced by two light bars (Fig. \ref{fig3}F) for microrheology of polymers and  manipulation tasks of cells. Rheological properties of polymer networks can be inferred from its deformation parameter $D_f = (L-B)/(L+B)$ and flow strain rates in an extensional flow, where $L$ and $B$ are the length and breadth of the deformed aster respectively (Fig. \ref{fig3}G). We use the two-bar flow to stretch a  microtubule aster pre-formed with a circular light pattern (Fig. \ref{fig3}G).  Measurement of  strain rates  at the gap center  shows that the flow is quasi-steady since $t=80$ s, and the average value is $\partial u_x/\partial x\approx 0.0015$ s$^{-1}$ (Fig. \ref{fig3}G). Consequently the shear modulus of the microtubule aster is calculated to be $1 \times 10^{-7}$ Pa (Materials and Methods). This shows that our system can be potentially used in extensional rheology where viscoelastic properties of materials are deduced from their deformation profiles under straining \cite{squires10}, and also in single polymer dynamics \cite{schroeder18} where single polymers, such as DNA strands \cite{chu97}, need to be stretched \textit{in situ}.  To perform these functions, traditional microfluidics usually relies on  channel geometry to generate stretching flows \cite{foster21}, such as sculpting channel contraction \cite{ober13, mancuso17} or cross-slot geometries \cite{chu97}. However the former method cannot fix particles in the channel because the flow constantly carry them downstream; the cross-slot or Taylor's four-mill geometries \cite{taylor34} can fix the particle at an equilibrium point but it is difficult to move the particle to this point since it is mechanically unstable. In contrast,  active matter programming language can be flexibly implemented to stretch particles \textit{in situ}, by projecting  light patterns around  objects of interest.

We now show using active matter to separate an unconstrained cell cluster into isolated cells and measure intercellular adhesion. Separation of cells are central in cell sorting  applications \cite{shields15} and intercellular adhesion are important in understanding cell communication and development of tissues \cite{khalili15}. Systematic study of cell-cell detachment are usually conducted  through either atomic force microscopy (AFM) \cite{benoit00} or pulling two cells apart using two micropipettes \cite{chu04, maitre12}. Both methods require highly-skilled operators and also could harm the cells when constraining them with solid instruments.  In contrast our system can generate  extensional flows in the vicinity of cell clusters and separate them (Fig. \ref{fig3}H and Video S1). The cells used are Jurkat  cells, which are a human T cell line and known to express cell adhesion molecules such as integrins and CD2 receptors \cite{tozeren92, zhao22}. The local flow strain rates near the cells are around 0.004 s$^{-1}$  (Fig. \ref{fig3}G), and the application of flows for 500 -- 800 s leads to cell detachment. The detachment force applied by the flow is approximately $3\pi \mu a \Delta l \partial u_x/\partial x $ (Materials and Methods), where $a=10 \ \mu$m is the cell radius, and $\Delta l$ is  the center-to-center distance  of the cell pair (Fig. \ref{fig3}H). Using the average strain rate 0.004 s$^{-1}$, the detachment force is proportional to $\Delta l$ and  calculated in Fig. \ref{fig3}H. 




Cell clusters consisting of more than two cells can also be separated {\itshape in situ}, revealing more complex collective behaviors. Fig. \ref{fig3}I and Video S2 show separation of a four-cell cluster in the extensional flow. In addition to three cell-cell detachment events (1-2, 2-3 and 3-4 in Fig. \ref{fig3}I), we also observed two initially separated cells (2 and 4 in Fig. \ref{fig3}I) first coming into contact and then being detached, through sliding along the membrane of cell 3. This demonstrates that intercellular bonds can be dynamically formed and broken during separation of a cluster. Close to separation, $\Delta l$ is around 24 $\mu$m across all detachment events (Fig. \ref{fig3}H and \ref{fig3}I), which may be determined by the extent of intercellular bond elongation and cell deformation. We also see regimes of behaviors that might be consistent with extension and then rupture-- as have been discussed in the literature on the mechanical stretching and breaking of molecular bonds \cite{muller09}. The displacements of cells seem to go through three stages -- first increasing, then plateauing, and lastly rapidly increasing (Fig. \ref{fig3}H and \ref{fig3}I). This may suggest that the cell separation first undergoes an elastic stretching, followed by a rupture of molecular bonds, starting at the plateauing stage determined by the extension limit of molecular bonds, and in the end rapid detaching due to breakup of bonds. The detachment force and bond lifetime are around 0.16 pN and 600 s, respectively. Both of them are smaller than those measured by AFM, which are 30 pN and 1 s \cite{zhao22}. The reason may be that the rupture force decays exponentially with lifetime \cite{bell78,evans01,zhao22}. New applications like these can be motivated because our system allows for flexible sculpting of local flow fields near particles of interest, as opposed to traditional microfluidics which usually relies on pumps to generate a global flow throughout the entire channel.

\section*{Active mixing with rotating bars}

\begin{figure}[t]
	\centering
	\includegraphics[width=\linewidth,angle=0]{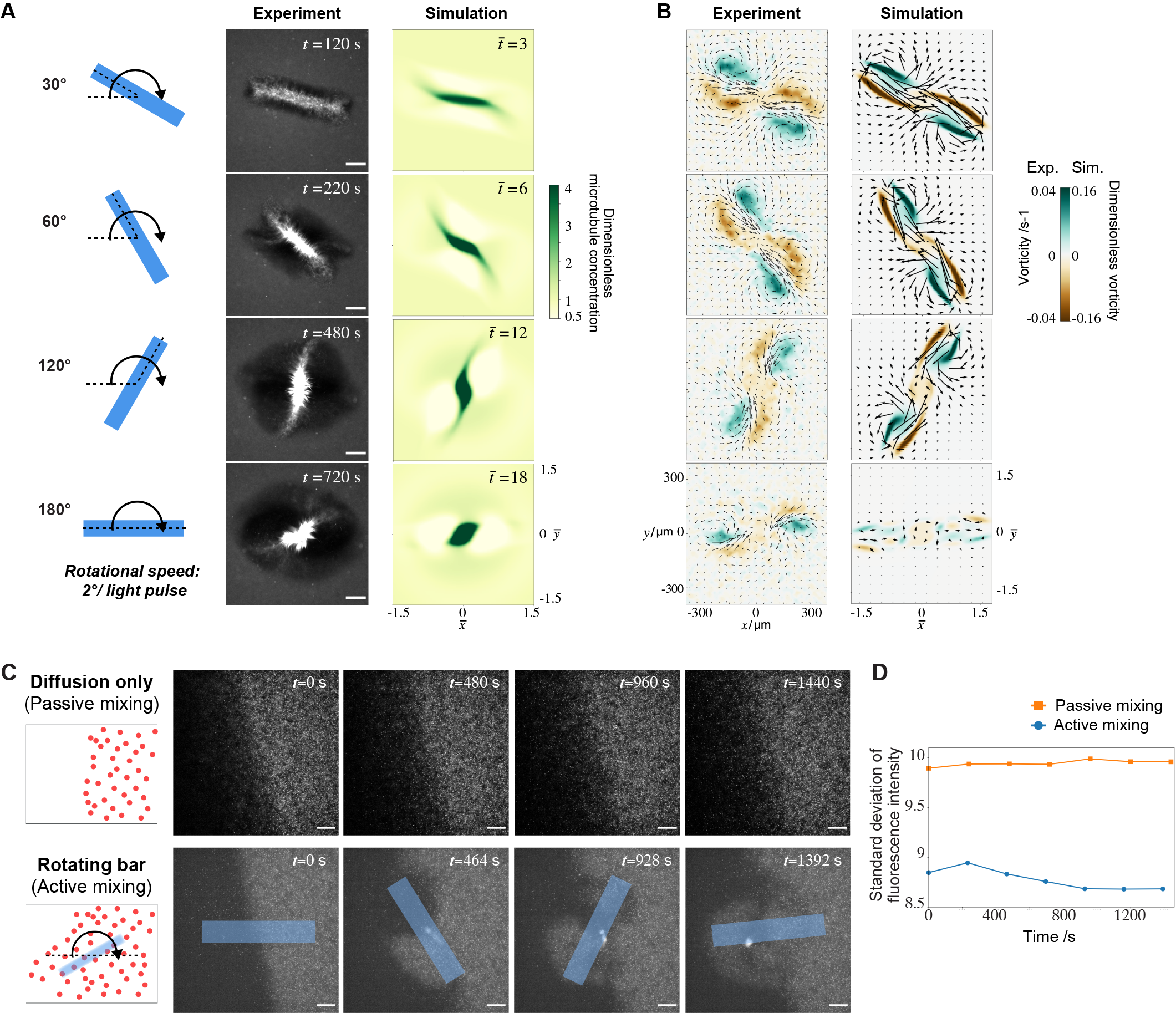}
	\caption{\textbf{Rotating light bars enables active mixing in microfluidics.}  \textbf{A}, Experimental and simulated images of microtubules under a rotating light bar. \textbf{B},  Measured and simulated flow fields in \textbf{A}. The measured flow fields are  averaged  over five experiments. \textbf{C}, A rotating bar can mix particles in the circular region swept by it. The controlled experiment on the top shows that the passive mixing  by diffusion is much weaker than the active mixing. \textbf{D}, Plot of the standard deviation of fluorescence intensity, which shows the non-uniformity of particle distributions, confirms that the rotating bar can actively mix the particles.  The rotational speed of light bars in \textbf{C} is 1$^\circ$/ light pulse. All scale bars are 100 $\mu$m.}
	\label{fig4}
\end{figure}

Our programming framework can also be extended to dynamic light patterns. Mixing at low Re is a major challenge in microfluidics because flows are laminar in this regime and molecular diffusion is slow \cite{stroock02}. We show that a rotating light bar can generate stirring flows to mix the circular region swept by it. Both the crosslinked microtubule gel and the flow field, consisting of four vortices,  rotate following the rotating light bar, as shown in Fig. \ref{fig4}A and \ref{fig4}B, which can also be predicted by our simulations. As the light bar rotates, the microtubule gel dynamically forms and self-contracts into a core. As the light bar is shed on a new location from one pulse to the next, two plumes of newly-crosslinked microtubules are formed at the  ends of the gel, which transport both mass and momentum of microtubules to the network core, as well as rotates the more densely-crosslinked network center (Fig. \ref{fig4}A). After one cycle, the circular region swept by the light bar will be mixed by the rotating vortices  (Fig. \ref{fig4}B). 

We qualitatively  test the mixing efficiency of the active mixer made from a rotating bar  on fluorescent particles, and find its mixing efficiency significantly higher than  the passive mixing by diffusion  (Fig. \ref{fig4}C and \ref{fig4}D). The active mixing is mainly accomplished by three flow effects at different length scales. At the length scale of the light bar, $w$, the rotating bar can push particles from one side of the interface to the other. At the length scale of the vortices,  $w/2$, each region occupied by one of the four vortices is mixed by the vortex. At the length scale of the thermal motions of  the active microtubules, the small-scale chaotic flows can also help randomly mix the particles \cite{wu17}.  Mixing is notoriously difficult in microfluidics, where current  techniques usually require  textured surfaces \cite{stroock02} or external energy input \cite{lee11} to stir  the flow. The stirring flows then need to pass through a long channel to ensure enough time for mixing. Our setup provides a promising method for stirring regions locally without passing through a long channel, because the dynamic optical signals can generate transient mixing flow patterns in a fixed position.

\begin{figure}[H]
	\centering
	\includegraphics[width=1\linewidth,angle=0]{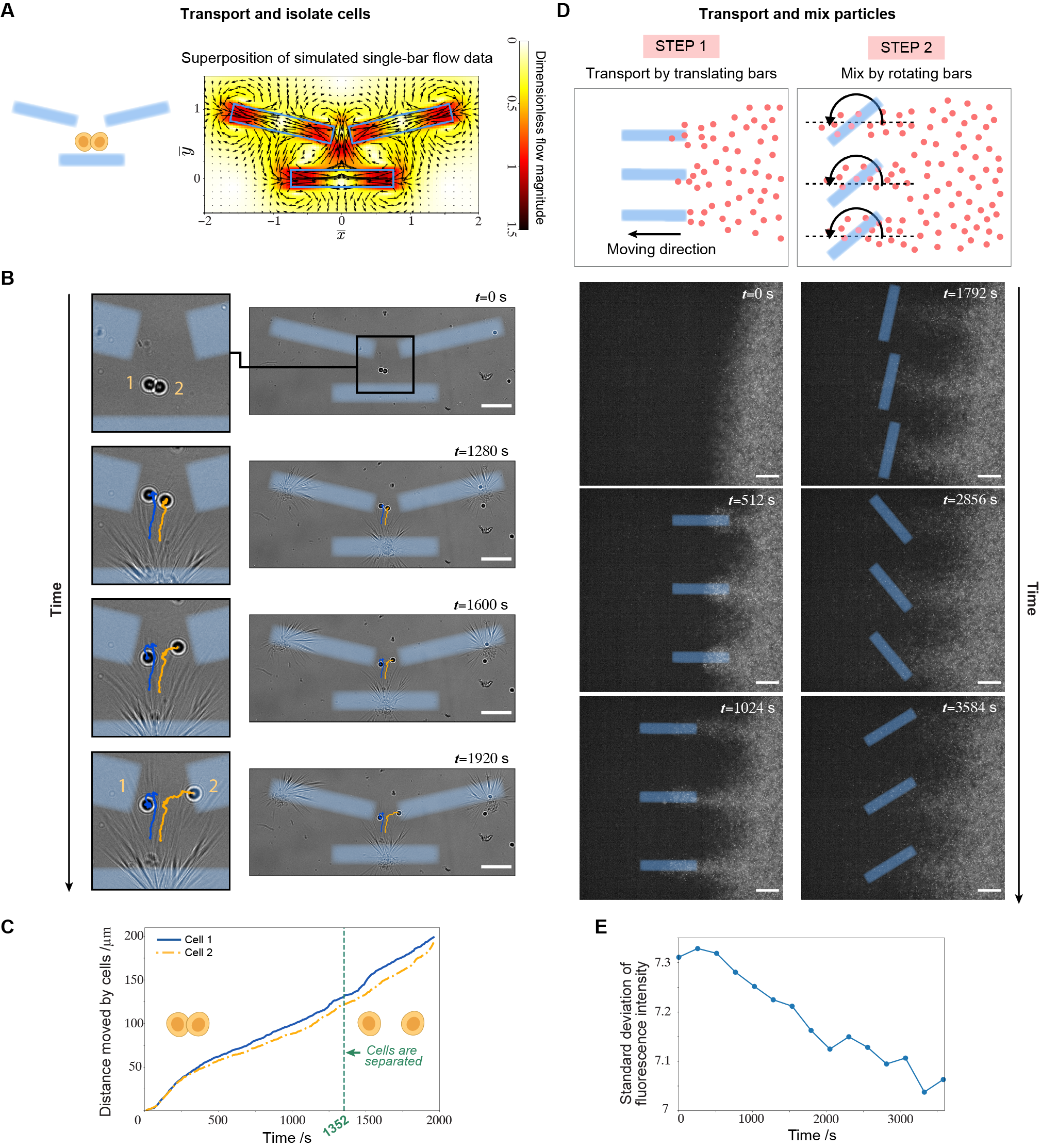}
	\caption{\textbf{ Programming modules can be additively assembled in space and time to   streamline multi-step tasks.}  \textbf{A}, Combination of  transporting and stretching modules can be designed from superposition of single-bar flow data. \textbf{B},  Experimental images of transporting and isolating cells at the same time. \textbf{C}, The distance traveled by the cells over time. \textbf{D}, Experimental images of  transporting and mixing particles by sequencing dynamic light signals in time. \textbf{E}, The standard deviation of fluorescence intensity over time. All scale bars are 100 $\mu$m.}
	\label{fig5}
\end{figure}


\section*{Superposition of primitives to construct multi-step flow programs}

Superposition enables assembly of programming modules in space and time to streamline multi-step tasks. We first show a module combination to transport and separate cells at the same time (Fig. 5A). The cells are propelled by the outflow generated by the bottom bar, and isolated by the two bars at the top. The flow fields can be easily predicted from linear superposition of simulated single-bar flow data (Fig. 5A), which can also be used for adjustment of the bar positions and inclinations. For example, we design the two top bars to tilt towards the bottom to facilitate the cell transport. Experiments verify that the module combination can transport and isolate cells at the same time (Fig. 5B and 5C).

We also sequence two dynamic light signals in time to transport and mix particles at a larger scale (Fig. 5D). We first transport the particles from right to left using three translating bars. The inflows of  light bars can attract particles that form into three stripes following the bars' movement. We then use three rotating bars to mix the stripes of particles, which disperse in the  region originally devoid of particles. The decrease of the standard deviation of fluorescent intensity also verifies the mixing effects of the combined modules (Fig. 5D).

\section*{Discussion}

In this paper, we harness biological arrive matter for engineering applications, opening the way for a broad range of active-matter-powered devices and providing an entirely new paradigm for flow control and materials science. Active materials generate force at molecular scales and can be exploited as a hardware in technology \cite{omar19}, but systematic control of active matter has been a challenge. Here, we demonstrate a novel  pathway by modulating the active-matter motions with a programmable external field, light. Furthermore, the dynamics of active fluids can  be quantitatively predicted by a  continuum model and model-driven design and optimization of flow fields are realized.

Linear superposition of local coherent flows  allows for methodical design of complex flow fields and scale-up of our system.  While active fluids are generally thought to be chaotic and difficult to control, we find that the nonlinear interactions of active networks decays as a power law with their separation length. This leads to a programming principle where we first construct local regions of organized flows and then isolate them above a critical spacing so that the networks are weakly-interacting and the local flow fields can be additively assembled in the entire channel. Our modular programming strategies also provide new theoretical insights into how to maintain order and perform work in active fluids.  Our theory gives a general framework for microtubule-motor-solvent systems, which may be extended to study other out-of-equilibrium structures of cytoskeleton, such as mitotic spindles. All experiments in this paper are controlled to be at similar conditions and all simulations share the same physical parameters (Supplemental Information), demonstrating the versatility of our model.

We envisage our platform can be developed towards building a universal single device that can integrate and automate diverse micron-scale transport tasks in a single channel, and will motivate numerous applications in chemistry and biology. Further improvements to our system include expanding the toolbox of basic flow units for more adaptable flow design, by incorporating flow fields generated from different light shapes, such as polygons, ellipses and concave shapes.  Our setup could also be  developed into a closed-loop control system by integrating a computer-vision  program to analyze the images of the channel, and a decision-making program to compute the optical  input based on real-time feedback and task goals. We note that an electrically-programmable microfluidic device has been proposed recently using artificial cilia to stir flows near the chamber walls  \cite{wang22}, while our setup is more efficient in driving flows near the center. Different programmable-control mechanisms might complement each other towards a fully-automated future of microfluidics.



\begin{acknowledgements}
	We thank Jerry Wang, David Larios and John Brady for fruitful discussions. We are also grateful to Howard A. Stone, Zev Gartner, Allen Liu and Alec Louren\c{c}o for valuable advice on the manuscript. We thank Inna-Marie Strazhnik for preparing the figures and proofreading the text. This  work is funded by the Packard Foundation, Moore Foundation, the National Institute of Health and  the Donna and Benjamin M. Rosen Bioengineering Center. F.Y.  acknowledges support from the BBE Divisional Fellowship at Caltech.
\end{acknowledgements}

\section*{ Materials and Methods}

\textbf{Kinesin Purification, Microtubule Polymerization}. Kinesin purification, microtubule polymerization and chamber construction were described in previous work \cite{ross19}. In short, we constructed and purified two kinesin K401 with light-induced hetero-dimer system of iLID and SspB-micro: K401-iLID and K401-micro. For protein expression, we transformed the plasmids into BL21pLysS cells and induced the cells with IPTG. For protein purification, we lysed the cells and used Ni-NTA agarose resin to pick up His tags that were provided by the base plasmids. MBP domain was used and subsequent cleaved off in K401-micro expression to ensure the micro domain remains fully functional during expression. Tubulin was polymerized with the non-hydrolysable GTP analog GMP-CPP \cite{georgoulia12}. Labeled and unlabeled tubulin were palleted and then incubated at 37$^\circ$C to form GMP-CPP stabilized microtubulues. We then characterized the microtubule length distribution by immobilizing them onto cover glass surface using poly-L-lysine.

\textbf{Flow chamber treatment and construction}.
The chambers were made from microscope slides and cover slips that were passivated against non-specific protein binding with a hydrophilic acrylamide coating \cite{lau09}. In brief, microscope slides and cover glass are first cleaned by sonication in 2\% Hellmanex III solution for 15 minutes. Excess Hellmanex III are then washed out with ddH\textsubscript{2}O and then ethanol sonication. The glass is then incubated overnight in 0.1M HCl to remove any trace metal and finished in 0.1M KOH sonication. After cleaning and etching, the glass us unnersed in a silanizing solution of 98.5\% ethanol, 1\% acetic acid, and 0.5\% 3-(Trimethoxysilyl)propylmethacrylate for 10-15 minutes. After rinsing, the glass is baked at 110$^\circ$ C for 30 minutes. The glass is than immersed overnight in a degassed 2\% acrlylamide solution with 0.035\% TEMED and 3mM ammonium persulfate. The glass is rinsed in ddH\textsubscript{2}O and air dried just before use. A flow cell made with pre-cut parafilm was used to seal between the microscope slides and coverslips making a channel that is about 70 $\mu$m in height. After the addition of reaction mixture, the flow cells were sealed with dental silicone polymer.

\textbf{Energy Mixture and Reaction Mixture}.
An energy mix consisting an energy source (ATP), a crowding agent (glycerol), a surface passivating reagent (pluronic acid), oxygen scavengers (glucose oxidase, glucose, catalase, Trolox, and DTT) and ATP-recycling reagents was made  on ice prior to combining the motor proteins and microtubules. After equilibrating the energy mix to ambient temperature, K401-micro, K401-iLID and microtubules were combined with the energy mix into a reaction mix. Concentrations for protein monomers for the K401-micro and K401-iLID constructs were 1 $\mu$M, and for microtubules, 1.5 – 2.5 $\mu$M. To minimize unintended light activation and non-specific protein binding, the sample was prepared under dark-room conditions with filters to block wavelengths below 580 nm. 
For all experiments conducted for this study, the reaction mixture consisted of 59.2 mM K-PIPES pH 6.1, 4.7 mM
MgCl2, 3.2 mM potassium chloride, 2.6 mM potassium phosphate, 0.74 mM EGTA, 1.4 mM MgATP
(Sigma A9187), 10\% glycerol, 0.50 mg/mL pluronic F-127 (Sigma P2443), 0.22 mg/ml glucose oxidase (Sigma G2133), 3.2 mg/mL glucose, 0.038 mg/mL catalase (Sigma C40), 5.4 mM DTT, 2.0 mM Trolox (Sigma 238813), 0.026 units/$\mu$l pyruvate kinase/lactic dehydrogenase (Sigma P0294), and 26.6 mM phosphoenolpyruvic acid. K401-micro, and K401-iLID were both diluted with a 1:2 ratio with 2$\mu$L of M2B pH 6.1 (80 mM K-PIPES pH 6.1, 1 mM EGTA, 2 mM MgCl2). Microtubules were diluted with a 1:7 ratio with 7$\mu$L of DTT M2B pH 6.1 (45 $\mu$L of M2B pH 6.1 with 1$\mu$L 250 mM DTT and 333.4 mg/$\mu$L glucose). The reaction mix was then aged in the flowcell for 120 - 180 minutes before light activation and data acquisition.

\textbf{Tracer Beads Preparation}.
To visualize the fluid dynamics of our system, we used 1 $\mu$m polystyrene beads as tracer particles. The particles were incubated overnight in M2B pH6.8 buffer with 50 mg/ml pluronic acid. The beads were then washed and palleted at 1000xg for 2 minutes and resuspended in M2B pH 6.8 before adding them into the reaction mix.

\textbf{Fluorescent Beads Preparation}.
Fluorescent particles were used to demonstrate mixing and transport capability of our system. We used 0.5 $\mu$m polystyrene beads that are dyed with highly hydrophobic dyes. The particles were incubated in M2B pH6.8 buffer iwth 50 mg/mL pluronic acid. The beads were then washed and palleted at 1000xg for 2 minutes and resuspended in M2B pH 6.8 buffer before adding them into the reaction mix.

\textbf{Cell Culture}
The cells used in the transport study (Jurkat cells; ATCC TIB-1522) were cultured in medium composed of high-glucose RPMI 1640 (Life Technologies, Carlsbad, CA, USA) and 10\% fetal bovine serum (FBS; Qualified, Life Technologies, Carlsbad, CA, USA). Jurkat cells were cultured to maintain cell density between $1 \times 10^{5}$ and $3 \times 10^{6}$ cells/mL. Before loading the cells in the aster mix, the cells were thoroughly washed with M2B buffer 6.8 (previously described in Materials and Methods). Cells culture were first centrifuged at 300x at 5 minutes to remove the culture media. Then washed twice with M2B 6.8 at 300x at 5 minutes to remove any remaining culture media and salts. Subsequently, cells suspended in M2B 6.8 were introduced into the microtubule buffer to attain the desired cell density. As an example, for a 5mL culture with a density of $3 \times 10^{6}$ cells/mL, the typical protocol would involve suspending the cells in 1 mL of M2B 6.8, of which 10 $\mu$L would be utilized in every 45 $\mu$L of the microtubule buffers.

\textbf{Design and Implementation of Different Bar Patterns}.
We custom fitted an epi-illuminated pattern projector onto our microscope. The size of the projection field is 800 x 1280 pixels.  Matrices containing coordinates of bars were first computed in Python and then converted to greyscale and eventually saved into tiff formats. Tiff image sequences were then processed by custom micro-manager script. 
The scripts can be found at https://github.com/fy26/ActiveMatter.

\textbf{Data Acquisition and Projection of Patterns}.
All experiments were performed with an automated widefield epifluorescence microscope with custom epi-illuminated projector and LED gated transmitted light discussed in our previous work \cite{ross19}. All samples were imaged at 10X magnification. Image sequences were captured using a Nikon TI2 controlled with micro-manager. Images of the fluorescent microtubules (cy5) and tracer particles (brightfield) were also acquired every 8 seconds. Bar patterns are projected onto the image plane every 8 seconds with a brief 200 milliseconds flash of a 2.4 mW/mm2 activation light from a 470 nm LED. The durations of activation lights were empirically determined by gradually increase or decrease the activation time based on the activity of the samples.

\textbf{Particle Image Velocimetry (PIV)}. 
PIV was performed on the images of  tracer beads using PIVlab \cite{thielicke14, thielicke21} to extract the solvent flow fields. 

\textbf{Derivation of the general principle for linear superposition}.
The generalized Stokes equation for the solvent flow is $ \bs{f}+  \mu \nabla^2 \bs{u} -\bs{\nabla} p  =0$, where $\bs{f}$ is the body force applied by external fields or sources, for example, in our system the body force is from the active force generated by the crosslinked microtubules and motors. We now construct a general principle for linear superposition of flows induced by $n$ different force-generating sources. We denote the force  applied on the fluid from the source $i$, in the absence of other sources, as $\bs{f}_i$, and the resultant flow and pressure field as $\bs{u}_i$ and $p_i$, respectively. Similarly, the body force, flow field and pressure field in the presence of all the $n$ sources are denoted by $\bs{f}_t$, $\bs{u}_t$ and $p_t$, respectively. To establish a linear regime of fluid flows driven by different sources, we require $\bs{u}_t = \sum_{i} \bs{u}_i$ and $p_t = \sum_{i} p_i$, which can be substituted into the Stokes equation and yield $\bs{f}_t = \sum_{i} \bs{f}_i$. This result directly comes from the linearity of the Stokes equation in $\bs{u}$ and $p$, however, it is not trivial because $\bs{f}_i$ can be dependent on the flow velocity $\bs{u}$ and the formula $\bs{f}_t = \sum_{i} \bs{f}_i$ does not always hold. In our system, the force-generating sources are the microtubule networks and the force induced by a single network $i$, in the absence of all other networks, can be expressed as $\bs{f}_i = \gamma c_i (\bs{v}_i - \bs{u}_i)$, where $\gamma$ is the drag coefficient of the microtubule and the  solvent, $c$ and $\bs{v}$ are the density and velocity of the microtubule network, respectively. Therefore, in the presence of $n$ networks, the general principle $\bs{f}_t = \sum_{i} \bs{f}_i$ together with the additional  linear relationships,  $c_t = \sum_{i} c_i$  and $\bs{v}_t = \sum_{i} \bs{v}_i$, requires that $\sum_{i} c_i \cdot \sum_{j} \bs{v}_j =  \sum_i c_i \bs{v}_i$ and $\sum_{i} c_i \cdot \sum_{j} \bs{u}_j =  \sum_{i} c_i \bs{u}_i $. The former requires that $c_i \bs{v}_j = \bs{0}$ when $i \neq j$, which is automatically satisfied as long as there are no  networks overlapping in the system. The latter requires that $c_i \bs{u}_j = \bs{0}$ for any $i \neq j$, which is the rule of linear superposition in a multiple-active-agent system. 

\textbf{Numerical Simulation}.  Finite difference method was used in numerical simulations with the central differencing scheme in space and the method of lines in time.   
The codes are written in Python and available at
https://github.com/fy26/ActiveMatter.

\textbf{Calculation of shear modulus $G$}. The steady-state value of $D_f$ depends on the Capillary number $Ca = \mu \frac{\partial u_x}{\partial x} r/G$ \cite{lac04, leal09, leonetti14}, where $r$ is the radius of the aster, via a linear relationship $D_f=A Ca$.  The value of coefficient $A$ is calculated to be $25/6$ for elastic capsules \cite{lac04, leonetti14},  and measured to be around 20  for viscoelastic drops \cite{leal09}. Here we choose $A = 10$ to estimate the shear modulus $G$ of the microtubule aster.  Additionally using measurements $\mu = 0.02$ Pa$\cdot$s \cite{gagnon20}, $\partial u_x /\partial x=0.0015$ s$^{-1}$,  $r=100 \ \mu$m, $L=120 \ \mu$m and $B=70 \ \mu$m, the shear modulus of the aster $G$ is calculated to be $1 \times 10^{-7}$ Pa.

\textbf{Calculation of detachment force on cells in an extensional flow}. The flow-induced friction $\boldsymbol{f}_p$ on a spherical particle translating in an unbounded fluid with velocity $\boldsymbol{v}$ is $\boldsymbol{f}_p=-6 \pi \mu a \boldsymbol{v}$, which is used to approximate the force on detaching cells. We denote the two attached cells by $a$ and $b$, and the unperturbed flow velocity at the two cell centers by $\boldsymbol{u}_a$ and $\boldsymbol{u}_b$, respectively. Then the cell pair moves at the same velocity $(\boldsymbol{u}_a + \boldsymbol{u}_b)/2$. The magnitude of the flow-induced force on each cell  is $f_p=3 \pi \mu a |\boldsymbol{u}_a - \boldsymbol{u}_b| = 3 \pi \mu a |\partial \boldsymbol{u}/\partial \boldsymbol{x}|\Delta l$. The detachment force on each cell has the same magnitude as $f_p$.

\bibliography{FlowControlref}

\pagebreak
\onecolumngrid

\begin{center}
\textbf{\large Supplemental Information}
\end{center}


\setcounter{equation}{0}
\setcounter{figure}{0}
\setcounter{table}{0}
\setcounter{section}{0}
\setcounter{page}{1}

\makeatletter
\renewcommand{\theequation}{S\arabic{equation}}
\renewcommand{\thefigure}{S\arabic{figure}}

\renewcommand{\bs}{\boldsymbol}
\renewcommand{\dbs}[1]{\dot{\boldsymbol{#1}}}
\renewcommand{\tn}{\widetilde{\bs{\nabla}}}
\renewcommand{\on}{\bs{\nabla}}
\renewcommand{\wb}[1]{\widetilde{\bs{{#1}}}}
\renewcommand{\bbb}[1]{\overline{\bs{#1}}}
\renewcommand{\obb}[1]{\hat{\overline{#1}}}
\renewcommand{\oa}[1]{\widetilde{\mathcal{#1}}}
\renewcommand{\rb}[1]{\bs{\mathcal{#1}}}
\renewcommand{\pf}[2]{\frac{\partial #1}{\partial #2}}
\renewcommand{\pfw}[2]{\frac{\partial \widetilde{#1}}{\partial \widetilde{#2}}}
\renewcommand{\pfm}[3]{\frac{\partial^#1 #2}{\partial #3^#1}}
\renewcommand{\pft}[2]{\frac{\partial^2 #1}{\partial #2^2}}
\renewcommand{\pfh}[2]{\frac{\partial^3 #1}{\partial #2^3}}
\renewcommand{\df}[2]{\frac{\mathrm{d} #1}{\mathrm{d}  #2}}
\renewcommand{\dfm}[3]{\frac{\mathrm{d} ^#1 #2}{\mathrm{d}  #3^#1}}
\renewcommand{\dft}[2]{\frac{\mathrm{d} ^2 #1}{\mathrm{d}  #2^2}}
\renewcommand{\md}{\mathrm{d}}
\renewcommand{\f}{p^\text{off}}
\renewcommand{\n}{p^\text{on}}
\renewcommand{\fw}{P^\text{off}}
\renewcommand{\nw}{P^\text{on}}
\renewcommand{\m}{p^\text{m}}
\renewcommand{\pd}{p^\text{d}}
\renewcommand{\mw}{P^\text{m}}
\renewcommand{\dw}{P^\text{d}}
\renewcommand{\ab}[1]{\langle #1 \rangle}
\renewcommand{\abp}[1]{\langle #1 \rangle_{\bs{p}}}
\renewcommand{\abpp}[1]{\langle #1 \rangle_{\bs{pp}}}
\renewcommand{\re}{\rho_\text{eq}}
\renewcommand{\R}{R_\text{eq}}
\renewcommand{\w}[1]{\widetilde{#1}}
\renewcommand{\pe}{\mathrm{Pe}}

\renewcommand{\nb}{\bs{\nabla}_{\bs{x}}}
\renewcommand{\np}{\bs{\nabla}_{\bs{p}}}
\renewcommand{\Dt}[1]{\frac{\mathrm{D} #1}{\mathrm{D} t}}

\renewcommand{\ol}[1]{\overline{#1}}

\newcommand{\Pe}{\text{Pe}}
\newcommand{\ob}[1]{\overline{\boldsymbol{{#1}}}}
\newcommand{\obn}{\ob{\nabla}}


\section{Comparison of flow magnitudes between superposition and experiments}

\begin{figure}[H]
	\centering
	\includegraphics[width=0.9\linewidth,angle=0]{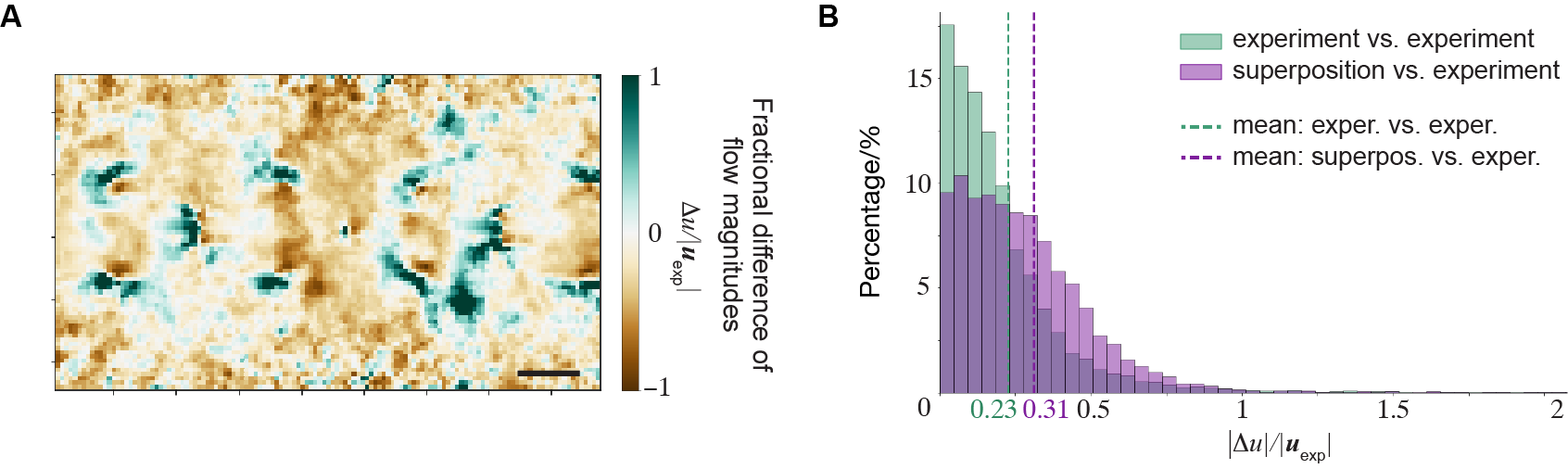}
	\caption{\textbf{A}, Comparison of flow magnitudes between two-bar experiments and  superposition of 2 single-bar flows.  The heatmap represents the fractional difference between measured and superposed flow magnitudes $\Delta u/|\bs{u}_{\mbox{exp}}|$, where $\Delta u = |\bs{u}_{\mbox{sup}}| -|\bs{u}_{\mbox{exp}}|$. \textbf{B}, Distributions of discrepancy in \textbf{a} and experimental variations. This demonstrates that the superposition-induced discrepancy is relatively small, only increasing the mean of errors from 0.23 to 0.31.  All scale bars are 100 $\mu$m and all  flow fields are time-averaged from $t=160$ s to $400$ s.}
	\label{SMfig0}
\end{figure}

To quantify the errors introduced by superposition, we compare the flow magnitudes between the superposed flow of 2 single-bar flows and experimental measurement of 2-bar flows (Fig. \ref{SMfig0}A). The flow data are the same with Fig. 1B and 1C in the main text. We also calculate the experimental variations of flow magnitudes and compare their distribution with the superposition-induced error distribution, which is plotted in Fig. \ref{SMfig0}B. The results show that the errors induced by linear superposition are small, only increasing the mean of errors from 0.23 to 0.31.

\section{Coarse-grained model} \label{sec::model}
\subsection{Overview}

Active networks of microtubules and motors exhibit diverse, and sometimes seemingly contradictory, behaviors under different conditions \cite{marchetti13, needleman17}. For example, both contractile \cite{foster15, ross19} and extensile \cite{sanchez12} microtubule gels have been observed. Classified by symmetry, microtubule networks can exist in isotropic \cite{qu21}, polar \cite{dalton22} or nematic \cite{sanchez12, decamp15} states. The active contraction in our system is mainly driven by the number density difference of the dimer motors and microtubules, where the isotropic contractile forces are the central driving force. Indeed, such forces have been found to induce the active contraction of microtubule stripes \cite{foster15}. 

In the past two decades, various coarse-grained models for motor-microtubule systems have been derived based on thermodynamic principles, such as the active gel theory \cite{joanny07, prost15, julicher18}, or microscopic interactions \cite{ahmadi06, furthauer19, furthauer21}. However, successful comparison of models and experiments for both microtubules and the ambient fluid is still lacking. Past attempts at modeling  fluid flows in motor-microtubule systems are mainly through the boundary force model \cite{ross19, qu21}, by placing force singularities at the boundaries of the illuminated region. The singularities are assumed immobile and thus the dynamics of active microtubules is not captured with this approach. Here we construct a multiphase complex fluids model that can quantitatively predict the coupled dynamics of microtubules and solvent flows. 

Our system consists of three phases, namely the crosslinked microtubules,  the freely-moving microtubules and the solvent solution. Other chemical species in our model include  ATP and motor proteins. Under illumination, the monomer motor proteins will  dimerize into dimer motors, which can  be  further classified depending on whether they are bound on microtubules or not. The chemical reactions in our system  include the reversible dimerization of motors under light and  the reversible crosslinking of microtubules with dimer motors. These chemical reactions will enter our model through the transport equations in section \ref{SM::subs::cont}.

In terms of dynamics, our system is mainly driven by the active stresses generated in the crosslinked microtubules. The stress-strain response of the crosslinked microtubules is modeled as a viscoelastic gel.    The free microtubules are treated as passive particles and the solvent solution a Stokes flow.  The ATP and motor proteins are much smaller molecules and assumed to follow the solvent flow, with  the exception of the dimer motors bound on microtubules, which follow the crosslinked microtubules. Each of the three phases also exerts friction on the other two. The momentum equations will be given in section \ref{SM::subs::mom}

As discussed before, we find \textit{a posteriori} that the isotropic active stresses dominate in our system and the polar and nematic effects can be neglected. For simplicity we will not model the polarity distribution and its dynamical impacts in this paper.

\subsection{Continuity Equations} \label{SM::subs::cont}

\begin{figure}[t]
	\centering
	\includegraphics[width=0.55\linewidth,angle=0]{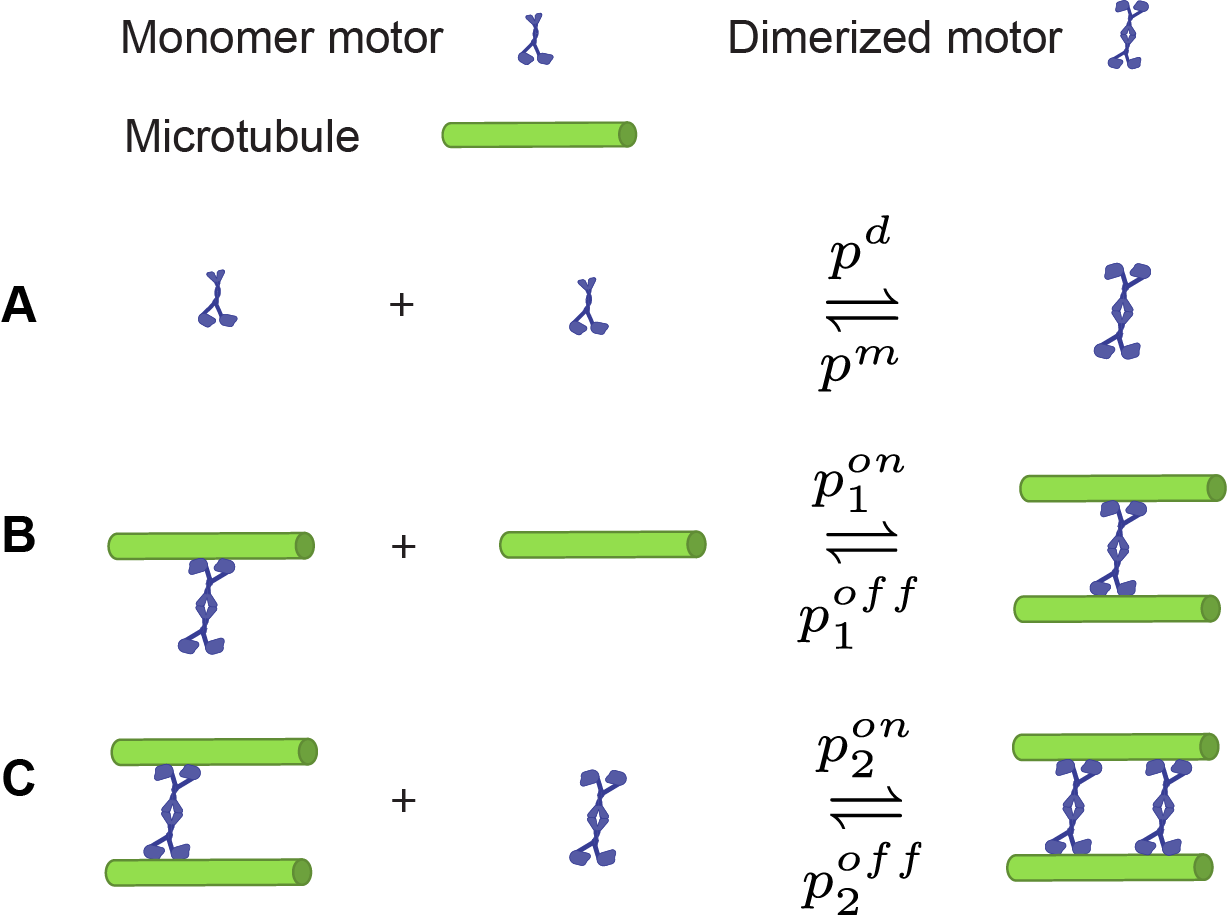}
	\caption{\textbf{Illustrations of chemcial reactions in the model}. \textbf{A}, Reversible dimerization of motors. \textbf{B}, Reversible crosslinking of a freely-moving microtubule $i$ and another microtubule $j$. The latter can be either free or pre-crosslinked. \textbf{C}, Reversible binding of a dimer motor on  two pre-crosslinked microtubules.}
	\label{SMfig1}
\end{figure}

The chemical reactions are sketched in Fig. \ref{SMfig1}. The motor proteins can reversibly dimerize and the transport equations are
\begin{equation}
\pf{m}{t} + \on \cdot (m \bs{u}) = 2 (p^m d_f - p^d m^2) + D_m \nabla^2 m \label{SM::m}
\end{equation}
where $m$ is the monomer motor concentration, $t$ is time, $d_f$ is the freely-moving dimer concentration, $\bs{u}$ is the solvent flow velocity,  $D_m$ is the diffusivity,  $p^m$ and $p^d(x,y,t)$ are the monomerization and dimerization rates, respectively. For a static rectangular light bar,   $p^d(x,y,t)$ is assumed to be
\begin{equation}
p^d(x,y,t) = \frac{p^d_0}{\left(1+\left(\frac{x-c_x}{r_x}\right)^{8} \right)\left(1+\left(\frac{y-c_y}{r_y}\right)^{8} \right)},
\end{equation}
which is approximately a rectangular function with its center at $(c_x, c_y)$, side lengths $2r_x$ and $2r_y$, and maximum value $p_0^d$. The exponent ``8" is empirically chosen. In general this exponent should be an even number and a larger exponent yields a sharper boundary of the rectangle. For a rotating light bar, one just needs to rotate the above expression in the $x-y$ plane at different times.

The transport equations for the crosslinked and freely-moving microtubules are
\begin{subeqnarray}
	\pf{c}{t}+\on \cdot \left( c\bs{v}\right) = p^{on}_1 c_f d_f - p^{off}_1 c,\\
	\pf{c_f}{t}+\on \cdot \left( c_f\bs{v}_f\right) = -p^{on}_1 c_f d_f + p^{off}_1  c + D_f\nabla^2 c_f. \label{SM::c}
\end{subeqnarray}
where  $D_f$ is the diffusivity of the freely-moving microtubules, $p^{on}_1$ and $p^{off}_1$ are the crosslinking and un-crosslinking rates, respectively, see Fig. \ref{SMfig1}B. The velocities of the crosslinked and freely-moving microtubules are denoted by $\bs{v}$ and $\bs{v}_f$, respectively.

The transport equations for the freely-moving and the bound  dimer motors are
\begin{subeqnarray}
	\pf{d_f}{t} + \on \cdot \left( d_f \bs{u}\right) = -p^m d_f + p^d m^2 - p_1^{on}c_fd_f-p_2^{on}cd_f+p^{off}_2d_b+D_d\nabla^2 d_f\\
	\pf{d_b}{t}+\on \cdot \left( d_b \bs{v}\right) =  p_1^{on}c_fd_f + p_2^{on}cd_f - p^{off}_2d_b, \label{SM::d}
\end{subeqnarray}
where $d_b$ is the bound dimer concentration,  $p_2^{on}$ is the rate of the dimer motors binding on the crosslinked microtubules, $D_d$ is the diffusivity of the free dimers, $p^{off}_2$ is the unbinding rate of a dimer motor without uncrosslinking the two microtubules, see Fig. \ref{SMfig1}C.  All freely-moving motors are assumed to follow the solvent flow and the bound motors follow the crosslinked gel.

The transport equation for the ATP is
\begin{equation}
\pf{c_A}{t} + \on \cdot(c_A \bs{u}) = -k_Ad_bc_A +D_A\nabla^2c_A, \label{SM::ATPd}
\end{equation}
where $k_A$ is the consumption rate of the ATP by the bound dimer motors, and $D_A$ is the ATP diffusivity.
The continuity equation for the solvent flow is
\begin{equation}
\on \cdot \bs{u} = 0. \label{SM::contu}
\end{equation}

\subsection{Momentum Equations}  \label{SM::subs::mom}

We find  it  crucial to model the crosslinked and freely-moving microtubules as two separate phases. The former can be modeled as a self-contracting active gel and the latter passive particles. The solvent flow is assumed to be a Stokes flow. In this section, we list the momentum equations for each phase.

\subsubsection{Crosslinked microtubules as an active gel}

The crosslinked microtubules form an active gel, which is assumed to be viscoelastic. The momentum equation can be written as
\begin{equation}
\on \cdot \left(\bs{\sigma}_a + \bs{\sigma}_v+ \bs{\sigma}_{el} + \bs{\sigma}_{st}\right) + \bs{f}_{fl} + \bs{f}_f = \bs{0}, \label{SM::moma}
\end{equation}
where $\bs{\sigma}_a$, $\bs{\sigma}_v$,  $\bs{\sigma}_{el}$ and $\bs{\sigma}_{st}$ are the active, viscous, elastic and steric stresses in the gel, $\bs{f}_{fl}$ and $\bs{f}_f$ are the frictions with the solvent and freely-moving microtubules, respectively. Their explicit expressions can be derived from microscopic interactions \cite{furthauer19,furthauer21}. The active contractile stress is assumed to be
\begin{equation}
\bs{\sigma}_a = \alpha c_A d_b c \bs{I}, \label{SM::sigmaa}
\end{equation}
where $c$, $c_A$ and  $d_b$ are the concentrations of the crosslinked microtubules, ATP and the bound dimer motors, respectively, $\alpha$ is a constant activity coefficient and $\bs{I}$ is the identity tensor. This active stress is derived to be $\bs{\sigma}'_a \propto c^2 \bs{I},$ in Ref \cite{furthauer21}, where the bound motor concentration is implicitly assumed to be uniform and does not appear in active stresses. In our experiments, the concentration difference of the dimer motors inside and outside of the illuminated regions is the key to form active gels. We can modify the theory in Ref \cite{furthauer21} by adding a prefactor $d_b/c$ in the active stresses, which is the number of bound motors per microtubule. In another word, the expression (\ref{SM::sigmaa}) can be derived using the same methods in Ref \cite{furthauer21} by taking the motor concentration into account and assuming the activity is proportional to the ATP concentration.

The viscous stress is 
\begin{equation}
\bs{\sigma}_v=\eta d_b c \left( \on \bs{v} + \on \bs{v} ^T\right), \label{SM::sigmav}
\end{equation}
where $\bs{v}$ is the velocity of the crosslinked microtubules and $\eta$ is the viscosity coefficient. This viscous stress in Ref \cite{furthauer21} is $\bs{\sigma}'_v\propto c^2 \left( \on \bs{v} + \on \bs{v} ^T\right)$. In our model the coefficient is changed from $c^2$ to $d_bc$ to incorporate the effects of the motor concentrations. 

The elastic stress $\bs{\sigma}_{el}$ is assumed to follow the Oldroyd-Maxwell model with a long relaxation time \cite{alves21},
\begin{equation}
\Dt{\bs{\sigma}_{el}}-\left[ \bs{\sigma}_{el} \cdot \on \bs{v} + \on \bs{v}^T \cdot \bs{\sigma}_{el}  \right] = \eta_{el} d_b c \left( \on \bs{v} + \on \bs{v} ^T\right) - p^{off}_1 \bs{\sigma}_{el},
\end{equation}
where  $\eta_{el}$ is the elasticity coefficient and $\mathrm{D}/\mathrm{D}t$ is the material derivative. The steric stress is 
\begin{equation}
\bs{\sigma}_{st} =- \xi c^2 \bs{I},
\end{equation}
with a constant coefficient $\xi$.

The friction with freely-moving microtubules is 
\begin{equation}
\bs{f}_f = \beta cc_f(\bs{v}_f-\bs{v}),
\end{equation} 
where $\beta$ is the friction coefficient. The friction with the solvent flow is 
\begin{equation}
\bs{f}_{fl} = \gamma c(\bs{u}-\bs{v}) \label{SM::fl}
\end{equation}
with the drag coefficient $\gamma$. In general the hydrodynamic drag coefficient depends on the fiber orientation \cite{leal07}. In this paper we neglect the polarity in both active stresses and  hydrodynamic friction.

\subsubsection{Freely-moving microtubules}
The freely-moving microtubules are passive particles that only experience frictions from the gel and the ambient fluid. The force balance is $\beta c_fc\left(\bs{v}-\bs{v}_f \right) + \gamma c_f (\bs{u}-\bs{v}_f)=\bs{0}$, and their velocity is therefore 
\begin{equation}
\bs{v}_f = \frac{\beta c\bs{v} + \gamma \bs{u}} { \beta c + \gamma}. \label{SM::vf}
\end{equation}

\subsubsection{Solvent flows}

The flow cells used in the experiments are typical Hele-Shaw cells,  the horizontal dimensions ($x-$ and $y-$direction) of which greatly exceed their vertical dimension ($z$-direction). We now derive the averaged flow equation in the $xy-$plane. 
The three-dimensional momentum equation is
\begin{equation}
-\tn \w{\Pi} +\mu \w{\nabla}^2 \wb{u}+\gamma \w{c}  \left(\wb{v}-\wb{u}\right) + \gamma \w{c}_f  \left(\wb{v}_f-\wb{u}\right)=0, \label{SM::stokes}
\end{equation}
where $\w{\Pi}$ and $\mu$ are the fluid pressure and viscosity, respectively. We use a tilde ``$\sim$" on top to indicate the variable is a function of $(x, y, z)$ and if the variable is a vector, it is a three-dimensional vector, e.g., $\w{\Pi} = \w{\Pi}(x,y,z)$ and $\wb{u}(x,y,z) = \left(\w{u}_x(x,y,z), \w{u}_y(x,y,z), \w{u}_z(x,y,z)\right)$. Since the vertical length scale is much smaller than horizontal length scales, the classical lubrication theory \cite{leal07} shows that $\w{\Pi}(x, y, z) \approx \Pi(x, y)$, i.e., the fluid pressure is constant along $z$-direction, and $\w{u}_z\approx 0$. Substituting these two results into equation (\ref{SM::stokes}) yields $\w{v}_z\approx 0$ and $\w{v}_{f,z}\approx 0$. Furthermore, the $x-$ and $y-$components of equation (\ref{SM::stokes}) can be approximated by 
\begin{equation}
-\on \Pi +\mu \pft{\wb{u}}{z} +\gamma \w{c}  \left(\wb{v}-\wb{u}\right) + \gamma\w{c}_f  \left(\wb{v}_f-\wb{u}\right)=0, \label{SM::3dstokes}
\end{equation}
where all vectors only have $x-$ and $y-$components, i.e., $\wb{u} = \left(\w{u}_x(x,y,z), \w{u}_y(x,y,z)\right)$,  $\wb{v} = \left(\w{v}_x(x,y,z), \w{v}_y(x,y,z)\right)$ and  $\wb{v}_f = \left(\w{v}_{f,x}(x,y,z), \w{v}_{f,y}(x,y,z)\right)$. Therefore, the analytical solution of $\wb{u}$ requires knowledge of the three-dimensional distribution of $\w{c}$, $\wb{v}$ and $\wb{v}_f$. To derive a two-dimensional model, we further simplify equation (\ref{SM::3dstokes}) by replacing the frictions with their $z$-directional average, i.e.,
\begin{equation}
-\on \Pi +\mu \pft{\wb{u}}{z} +\gamma c  \left(\bs{v}-\bs{u}\right) + \gamma c_f  \left(\bs{v}_f-\bs{u}\right)=0, \label{SM::2dstokes}
\end{equation}
where $c(x,y)=h^{-1}\int_{0}^{h}\w{c}\md z$, $\bs{v}(x,y)=h^{-1}\int_{0}^{h}\wb{v}\md z$, $\bs{v}_f(x,y)=h^{-1}\int_{0}^{h}\wb{v}_f\md z$, and $\bs{u}(x,y)=h^{-1}\int_{0}^{h}\wb{u}\md z$ with the top and bottom walls of channel at $z=0$ and $z=h$, respectively. These notations are consistent with the rest of the paper. The solution of equation (\ref{SM::2dstokes}) is
\begin{equation}
\wb{u}=\frac{1}{2\mu}\left[\on \Pi -\gamma c\left(\bs{v}-\bs{u}\right) - \gamma c_f \left(\bs{v}_f-\bs{u}\right)\right]z(z-h). \label{SM::uf}
\end{equation}
Combining $\bs{u}=h^{-1}\int_{0}^{h}\wb{u}\md z$ and equations (\ref{SM::vf}) and (\ref{SM::uf}), we have the average two-dimensional flow velocity
\begin{equation}
\bs{u}=-\frac{h^2}{12\mu}\left[\on \Pi -\left(\gamma c + \gamma_f c_f\right)\left(\bs{v}-\bs{u}\right) \right], \label{SM::u}
\end{equation}
with $\gamma_f = \gamma \beta c /(\beta c+\gamma)$. The above result is a modified Darcy's law in a Hele-Shaw cell \cite{leal07} by incorporating the microtubule friction.

\subsection{Non-dimensionalization}

Using the initial microtubule concentration $c_0$, initial monomer motor concentration $m_0$, initial ATP concentration $c_{A0}$, and the typical length scale of the illuminated region $l$, we can non-dimensionalize the governing equations by
\begin{equation}
\begin{split}
\ol{c}=\frac{c}{c_0}, \quad \obn= l_0 \on, \quad \ol{m}=\frac{m}{m_0}, \quad \ol{d}_f=\frac{d_f}{m_0}, \quad \ol{d}_b=\frac{d_b}{m_0}, \quad \ol{c}_A=\frac{c_A}{c_{A0}} \\
\ol{t}=\frac{t}{t_0}, \quad  \ob{v}=\frac{\bs{v}}{v_0} , \quad  \ob{v}_f=\frac{\bs{v}_f}{v_0},  \quad \ob{u}=\frac{\bs{u}}{v_0},\quad \ob{\sigma}_{el} = \frac{\bs{\sigma}_{el}}{\sigma_0},\quad \ol{\Pi}=\frac{\Pi}{\Pi_0},
\end{split}
\end{equation}
where we use  overlines to denote dimensionless variables, $t_0$, $v_0$, $\sigma_0$ and $\Pi_0$ are  typical scales of time, velocity,  stress, and fluid pressure, respectively. By balancing the contractile stress (\ref{SM::sigmaa}) and the viscous stress (\ref{SM::sigmav}) in the active gel, we have $v_0 = \alpha c_{A0}l/\eta$. The  typical time scale follows as $t_0=l_0/v_0 = \eta /\alpha c_{A0}$. The typical stress scale can be obtained from equation (\ref{SM::sigmaa}), which is $\sigma_0 = \alpha c_{A0}c_0m_0$. From equation (\ref{SM::u}), we have $\Pi_0 = c_0\gamma v_0 l_0 =\alpha \gamma c_0 c_{A0} l_0^2/\eta $.

The dimensionless transport equations for the monomer (\ref{SM::m}) and dimer (\ref{SM::d}) motors are
\begin{subeqnarray}
	\pf{\ol{m}}{\ol{t}} + \obn \cdot (\ol{m} \ob{u}) = 2 (\ol{p}^m \ol{d}_f - \ol{p}^d \ol{m}^2) + \Pe_m^{-1} \ol{\nabla}^2 \ol{m}, \\
	\pf{\ol{d}_f}{\ol{t}} + \obn \cdot \left( \ol{d}_f \ob{u}\right) = -\ol{p}^m \ol{d}_f + \ol{p}^d \ol{m}^2 - \ol{p}_1^{on}\ol{c}_f\ol{d}_f-\ol{p}_2^{on}\ol{c}\ol{d}_f+\ol{p}^{off}_2\ol{d}_b+\Pe_d^{-1} \ol{\nabla}^2 \ol{d}_f,\\
	\pf{\ol{d}_b}{\ol{t}}+\obn \cdot \left( \ol{d}_b \ob{v}\right) =  \ol{p}_1^{on}\ol{c}_f\ol{d}_f + \ol{p}_2^{on}\ol{c}\ol{d}_f - \ol{p}^{off}_2\ol{d}_b, 
\end{subeqnarray}
where the dimensionless reaction coefficients are $\ol{p}^m =p^m t_0$, $\ol{p}^d =p^m m_0 t_0$, $\ol{p}^{on}_1 =p^{on}_1 c_0 t_0$, $\ol{p}^{on}_2 =p^{on}_2 c_0 t_0$ and $\ol{p}^{off}_2 =p^{off}_2 t_0$. We use $\Pe$ to denote the P\'{e}clet number and $\Pe_m = v_0 l /D_m$, $\Pe_d = v_0 l /D_d$.

The dimensionless transport equations for the microtubules (\ref{SM::c}) are
\begin{subeqnarray}
	\pf{\ol{c}}{\ol{t}}+\obn \cdot \left( \ol{c}\ob{v}\right) = \ol{p}^{on}_1 \ol{c}_f \ol{d}_f - \ol{p}^{off}_1 \ol{c},\\
	\pf{\ol{c}_f}{\ol{t}}+\on \cdot \left( \ol{c}_f\ob{v}_f\right) = -\ol{p}^{on}_1 \ol{c}_f \ol{d}_f + \ol{p}^{off}_1  \ol{c} + \Pe_f^{-1}\ol{\nabla}^2 \ol{c}_f
\end{subeqnarray}
with $\ol{p}^{off}_1 = p^{off}_1 t_0 $, $\Pe_f = v_0 l_0 /D_f$.  For ATP, equation (\ref{SM::ATPd}) becomes
\begin{equation}
\pf{\ol{c}_A}{\ol{t}}+\obn \cdot \left( \ol{c}\ob{u}\right) = -\ol{k}_A \ol{d}_b \ol{c}_A + Pe_A^{-1}\ol{\nabla}^2 \ol{c}_A,
\end{equation}
where $\ol{k}_A=k_A m_0 t_0$ and $\Pe_A = v_0 l_0 /D_A$.

The continuity equation for the solvent flow (\ref{SM::contu}) is
\begin{equation}
\obn \cdot \ob{u}=0.
\end{equation}

Using equations (\ref{SM::moma}-\ref{SM::vf}), the dimensionless momentum equations for the active gel are
\begin{equation}
\obn \cdot \left[\left(\ol{c}_A\ol{d}_b\ol{c} -\ol{\xi}\ol{c}^2 \right)\bs{I}+\obn \ob{v}+\obn \ob{v} ^T+ \ob{\sigma}_{el}\right] + \left(\ol{\gamma}_f\ol{c}_f + \ol{\gamma}\ol{c}\right)\left(\ob{u}-\ob{v}\right) = \bs{0}
\end{equation}
where $\ol{\xi} = \xi c_0 /\alpha c_{A0}m_0$, $\ol{\gamma}=\gamma l_0^2/m_0 \eta$,  $\ol{\gamma}_f=\gamma_f l_0^2/m_0 \eta$ and
\begin{equation}
\frac{\mathrm{D} \ob{\sigma}_{el}}{\mathrm{D} \ol{t}}-\left[ \ob{\sigma}_{el} \cdot \obn \ob{v} + \obn \ob{v}^T \cdot \ob{\sigma}_{el}  \right] = \ol{\eta}_{el} \ol{d}_b \ol{c} \left( \obn \ob{v} + \obn \ob{v} ^T\right) - \ol{p}^{off}_1 \ob{\sigma}_{el},
\end{equation}
with $\ol{\eta}_{el} = \eta_{el}/\alpha c_{A0}$.

The dimensionless velocity of the freely-moving microtubules (\ref{SM::vf}) is 
\begin{equation}
\ob{v}_f =\frac{\ol{\beta}\ol{c}\ob{v}+\ol{\gamma}\ob{u}}{\ol{\beta}\ol{c}+\ol{\gamma}},
\end{equation}
with $\ol{\beta}=\beta l_0^2 c_0 / m_0 \eta$. Note that only two of $\ol{\beta}$, $\ol{\gamma}$ and $\ol{\gamma}_f$ are independent, which are connected through $\ol{\gamma}_f =\ol{\gamma} \ol{\beta} \ol{c}/ (\ol{\beta} \ol{c} + \ol{\gamma})$.

The dimensionless solvent flow (\ref{SM::u}) is
\begin{equation}
\ob{u} = -\ol{\zeta}\left[ \obn \ol{\Pi}-\left(\ol{c}+\frac{\ol{\gamma}_f}{\ol{\gamma}}\ol{c}_f\right)\left(\ob{v}-\ob{u}\right)\right],
\end{equation}
with $\ol{\zeta} = h^2c_0 \gamma /12\mu$.

\subsection{Simulations}

\subsubsection{Physical parameters}
All experiments in the paper are controlled to be at similar conditions. Therefore, the same physical parameters are used in simulations throughout the paper, with one exception in Fig. 2C in the main text where $\ol{\zeta}$ is varied. The physical parameters are documented below:
\begin{eqnarray}
l_0 = 240 \ \mu \mbox{m}, \quad t_0 = 40 \ \mbox{s}, \quad p^d_0 = 15, \quad p^m = 12, \quad p^{on}_1 = 15, \quad p^{on}_2 = 0.75, \nonumber \\
p^{off}_1 = 10, \quad   p^{off}_2 = 20, \quad \Pe_f^{-1} =10^{-3}, \quad \Pe_m^{-1} = \Pe_d^{-1} = \Pe_A^{-1} = 10^{-2}, \nonumber\\
\ol{k}_A = 0.8,  \quad \ol{\xi}= 0.05, \quad \ol{\eta}_{el} = 1.5, \quad \ol{\gamma} = 0.045, \quad \ol{\beta}=0.45, \quad \ol{\zeta} = 20.5.
\end{eqnarray}
Additionally, the duration of each light pulse is 0.04 and the time interval between two light pulses is 0.2.

\subsubsection{Aspect ratios of single bars}

In the main text we show that the bar sizes do not affect the power law of  flow decay. Another important geometric factor is the aspect ratio. A previous study on flows generated with different aspect ratios (AR) \cite{qu21} shows that they can change the flow directions outside the light bar: when AR $\geq 0.5$, the flow is directed away from the light region, as opposed to when AR $\leq 0.25$, the flow goes into the light bar. 

\begin{figure}[t]
	\centering
	\includegraphics[width=0.9\linewidth,angle=0]{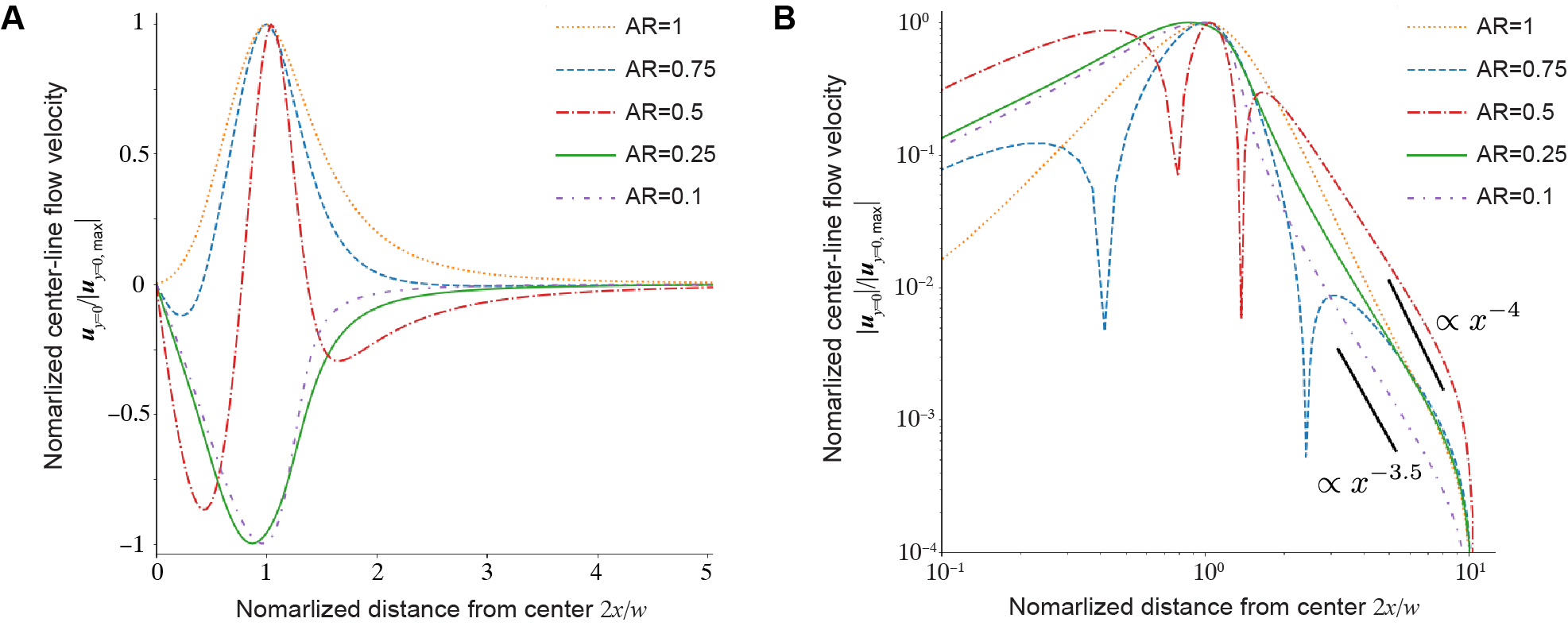}
	\caption{\textbf{A}. Simulated center-line flows induced by single bars with different aspect ratios (AR). Positive (negative) values represent the flow is directed away from (into) the light bar. \textbf{B}.  Absolute values of data in \textbf{A} in a log-log plot.}
	\label{SMfig2}
\end{figure}

Our simulations are in agreement with these observations (Fig. S3). To understand how the AR of a rectangular light bar affects scaling of  flow decay, we fix the bar width and simulate single-bar flows with different aspect ratios. The center-line flow magnitudes are documented in Fig. \ref{SMfig2}. Fig. \ref{SMfig2}A shows that aspect ratios can change flow directions. When AR$\leq 0.25$, the center-line flow is always pointing towards the bar center. When  AR$\geq 0.5$, the flow near the light bar edge is an outflow, pointing away from the bar. These findings are consistent with experiments \cite{qu21}. The log-log plot of flow data (Fig. \ref{SMfig2}B) shows that the flow outside the illumination region always decays following a power law, with the power ranging from -4 to -3.5.
Furthermore, our numerical results reveal that the flows outside the illuminated regions always decay in a power law, with the power ranging from -4 to -3.5, regardless of the aspect ratios (Fig. S3). 


\section{Optimization scheme for particle transport}

\begin{figure}[t]
	\centering
	\includegraphics[width=0.9\linewidth,angle=0]{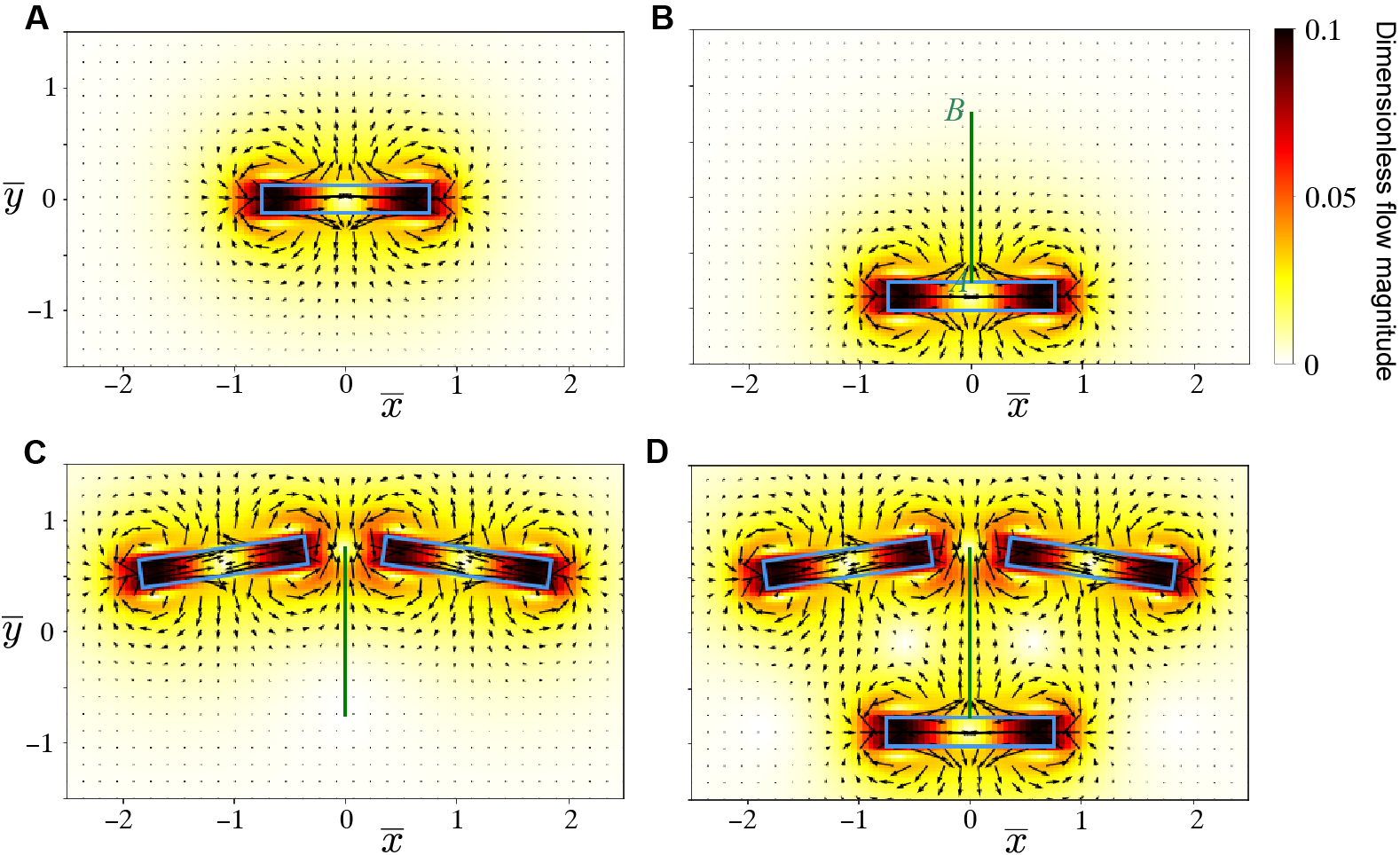}
	\caption{\textbf{A}, Simulated single-bar flow field. Optimal solutions for particle transport using \textbf{B}, $n = 1$, \textbf{C}, $n = 2$ and \textbf{D}, $n = 3$ light bars, and their corresponding flow fields. The objective is to move a particle along  the line segment AB (green). The outlines of light bars are plotted in blue. }
	\label{SMfig3}
\end{figure}

In this section we show how to optimize particle transport along a straight path using  linear transformation of single-bar flow data (Fig. \ref{SMfig3}A). To ensure that the particle move along a straight path, the line segment AB should coincide with the axis of symmetry in the composition of light patterns (Fig. \ref{fig3}B, main text): when using only one bar, the particle should be placed at the center line perpendicular to the longer sides of the light bar to avoid being absorbed into the aster; when using two bars, the particle should move along their axis of symmetry, with a constraint that the two bars are separate above a critical spacing $\overline{w}_c$ to avoid merging; when using three bars, the optimal solution is just a sum of the solutions using one and two bars. Therefore we only need to solve for the optimal solution using two bars.  This can be further simplified into finding the optimal location and orientation of a single bar to maximize $f$, and the second bar is just its reflection with respect to line AB. In this case, the objective function $f$ only depends on  three variables, the $\overline{x}-$ and $\overline{y}-$coordinates of the bar center, and its orientation angle. The optimal solution can be found by directly searching the maximum value over the three-dimensional optimization landscape using simulated single-bar flow data. The region $\overline{x}<\overline{w}_c+\overline{w}_y/2$, where $\overline{w}_y$ is the width of the shorter side, is prohibited due to the constraint $\overline{w}_g < \overline{w}_c$ (Fig. \ref{fig3}C, main text).

\subsection{Problem formulation}

Objective: Move a particle along a line segment  AB from $\ol{y}=-0.75$ to $0.75$ along the $\ol{y}-$axis and maximize its speed using $n$ bars. 

The objective function is defined as
\begin{equation}
f=\int_A^B \ob{u} \cdot \md \ob{x}. \label{SM::obj}
\end{equation}
The constraints are:
\begin{enumerate}
	\item The particle does not move into the aster.
	\item The spacing between bars should be above a critical length $\ol{w}_c$ to avoid interactions.
\end{enumerate}

\subsection{Solutions up to $n=3$}

To move the particle along a straight path, the axis of symmetry in the composition of light patterns should coincide with  the $\ol{y}-$axis.

\subsubsection{$n=1$}

There are two symmetry axes for a single rectangular bar (Fig. 1A, main text), and the fluid flow goes into the bar lengthwise and out of the bar breadthwise. To satisfy constraint 1, the light bar should be placed at one end of AB with its long side perpendicular to the $\ol{y}-$axis (Fig. \ref{SMfig3}B)

\subsubsection{$n=2$}

When there are two bars, the only way to construct a straight streamline, except for directly using the symmetry axes of single bars, is to make the two bars symmetric about  $\overline{y}$-axis. Therefore, the problem is reduced to finding the location and orientation of a single bar to maximize $f$, and the second bar is just its reflection with respect to $\overline{y}$-axis.

We denote the single-bar flow field in Fig. \ref{SMfig3}A as $\ob{u}_s$. We now calculate the flow field $\ob{u}$ induced by a single bar placed at $\ob{x}_s$ with an orientation angle $\psi \in [0,\pi)$. The rotation matrix $\bs{R}$ associated with $\psi$ is 
\begin{equation}
R(\psi) = 
\begin{pmatrix}
\cos \psi & -\sin \psi \\
\sin \psi & \cos \psi 
\end{pmatrix}.
\end{equation}
Through simple linear transformation of coordinates, the flow field $\ob{u}$ is just 

\begin{equation}
\ob{u}(\ob{x}) = R(\psi) \cdot \ob{u}_s \left(R(\psi)^T \cdot \left(\ob{x}- \ob{x}_s  \right) \right). \label{SM::op::u}
\end{equation}

The constraint 2 can be formulated as
\begin{equation}
\ol{x} - \frac{\ol{w}}{2} |\cos{\psi}|-\frac{\ol{w}_y}{2} \sin{\psi} > \frac{\ol{w}_c}{2}, \label{SM::OP::cons}
\end{equation}
where $\ol{w}$ and $\ol{w}_y$ are the length and breadth of the light bar.

The optimal solution  can be  found directly by searching for the maximum value of the objective function (\ref{SM::obj}) using (\ref{SM::op::u}) and the constraint (\ref{SM::OP::cons}), since it only depends on 3 variables. Due to symmetry, there is one optimal solution in each quadrant. We can further confine the computation domain to be in Quadrant II and the solution is $\ol{x}=-1.09$, $\ol{y}=0.62$ and $\psi = 8.2^\circ$ with $\ol{w}_c=0.325$. The overall optimal two-bar configuration is shown in Fig. \ref{SMfig3}C. 

\subsubsection{$n=3$}

Due to symmetry, the optimal solution for 3 bars can only be the linear superposition of the optimal solutions for $n=1$ and $n=2$, as shown in Fig. \ref{SMfig3}D. 

\bibliographystyle{unsrt}
\bibliography{selforg, FlowControlref}

\end{document}